\documentclass[useAMS,usenatbib]{mn2e}

\usepackage{amssymb}
\usepackage{amsfonts}
\usepackage{natbib}
\usepackage{epsfig}
\usepackage{mncite}

\newcommand{\de}{\mbox{d}}

\begin{document}

\title[Warp diffusion in accretion discs]{Warp diffusion in accretion discs: a
  numerical investigation}

\author[Lodato \&  Pringle] {Giuseppe Lodato$^{1,2}$ and J. E. Pringle$^{2,1}$\\
$^1$ Department of Physics and Astronomy, University of Leicester, Leicester,
LE1 7RH\\
$^2$ Institute of Astronomy, Madingley Road, Cambridge, CB3 0HA\\}

\maketitle

\begin{abstract}

  In this paper we explore numerically the evolution of a warped accretion
  disc. While previous analyses have concentrated on the case where the disc
  is thick enough that the warp propagates as a wave, we focus here on the
  opposite regime of a thin disc, where the warp evolves diffusively. By
  comparing the numerical results to a simple diffusion model, we are able to
  determine the diffusion coefficient of the warp, $\alpha_2$, as a function
  of the relevant disc parameters, such as its thickness and especially its
  viscosity. We find that while in general the disc behaviour is well
  reproduced by the diffusion model and for relatively large viscosities the
  warp diffusion is well described by the linear theory (in particular
  confirming that the warp diffusion coefficient is inversely proportional to
  viscosity), significant non-linear effects are present as the viscosity
  becomes smaller, but still dominates over wave-propagation effects. In
  particular, we find that the inverse dependence of the diffusion coefficient
  on viscosity breaks down at low viscosities, so that $\alpha_2$ never
  becomes larger than a saturation value $\alpha_{\rm max}$ of order unity.
  This can have major consequences in the evolution of systems where a warped
  disc is present. In particular, it affects the location of the warp radius
  in the Bardeen-Petterson effect and therefore the spin up (or spin down) of
  supermassive black holes in the nuclei of galaxies. Additionally, we also
  find that while the rate of warp diffusion does not depend significantly on
  the detailed viscosity formulation, the rate of internal precession
  generated by the warp is strongly affected by it. Such effects should be
  considered with care when modeling the evolution of warped discs. This
  emphasises the need to test the above results using different numerical
  schemes, and with higher resolution, in order to investigate the degree to
  which numerical simulations are able to provide accurate modeling of the
  complex fluid dynamics of warped discs.

\end{abstract}

\section{Introduction}

Many (perhaps most) astrophysically relevant accretion discs are
expected to be twisted or warped. The torques responsible for such
warping can be very different, including tidal effects of a companion
star in binary systems \citep{larwood96}, dynamical effects during the
formation of protostellar discs, general relativistic Lense-Thirring
precession around a spinning black hole
\citep{bardeen75,scheuer96,LP06}, and self-induced warping caused by
radiation pressure \citep{pringle96}.

Observationally, warps are found both in discs which are expected to be
relatively thick, such as the case of the hyperaccreting X-ray binary SS433
\citep{begelman06b} or, on the much less energetic side, the disc around the
young star KH 15D \citep{chiang04}. In some other cases, warps are also found
in very thin discs, in the nucleus of the AGN NGC 4258
\citep{herrnstein96,papaloizou98} or in the X-ray binary Her X-1
\citep{wijers99}.

The secular evolution of such warped discs deeply affects the structure of the
system. For example, the process of alignment between the spin of a black hole
and the angular momentum of the disc due to the Bardeen-Petterson effect
strongly depends on the effectiveness of warp diffusion in thin discs
\citep{LP06}. This can have major consquences on the spin up or spin down of
the black hole during an accretion event, having implications also on the
formation and early growth of supermassive black holes \citep{king06}.

While the linear (and mildly non-linear) theory of warp propagation has been
investigated extensively in the past
\citep{pappringle83,pringle92,paplin95,ogilvie99,ogilvie00} and has been
tested numerically in the case where the disc is thick enough that warp
propagation is wave-like \citep{nelson99,nelson00}, there is as yet no
numerical confirmation of the diffusive warp propagation expected in the thin
disc case, nor a full description of the non-linear effects that might arise
when the amplitude of the warp becomes large compared to viscosity.

We consider here the propagation of warps in thin Keplerian accretion discs in
the case when the disc is sufficiently viscous that the warp propagates in a
diffusive manner.

We start by defining the main disc properties. We consider a thin disc,
rotating with angular velocity $\Omega(R)$, with surface density $\Sigma(R)$
and angular momentum per unit area ${\bf L}(R)$. Here $R$ should be
interpreted as a `spherical' coordinate. The local direction of ${\bf L}$ can
be oriented anyhow in space, and the unit vector ${\bf l}(R)={\bf L}(R)/L(R)$
defines its direction. If the disc is rotating around a central point mass
$M$, then its rotation is Keplerian, with $\Omega=\sqrt{GM/R^3}$ and
$L(R)=\Sigma(R)\sqrt{GMR}$.

The disc is warped whenever the direction identified by ${\bf l}$ changes with
radius. We can characterize the warp amplitude with the dimensionless
parameter $\psi$, where
\begin{equation}
\psi=R\left|\frac{\partial{\bf l}(R)}{\partial R}\right|.
\end{equation}

The disc thickness is $H=c_{\rm s}/\Omega$, where $c_{\rm s}$ is the sound
speed, and is the scale over which density and pressure change in the local
$z$ direction. The disc aspect ratio is $H/R$, and we shall assume that
$H/R\ll 1$.

\citet{pappringle83} and \citet{pringle92} have introduced a simple
equation to describe the evolution of a warped disc, in the case where
the warp propagates diffusively in the disc. As shown by
\citet{paplin95}, diffusive behaviour occurs for isotropic viscosity
if the disc aspect ratio $H/R<\alpha$, where $\alpha$ is the viscosity
parameter commonly used as a measure of the $(R,\phi)$ (viscous)
stresses in the disc (Shakura \& Sunyaev, 1973).  In this case the
warp propagation can be approximately described by the equation
\citep{pringle92}

\begin{eqnarray}
\label{eq:pringle}
\nonumber\frac{\partial{\bf L}}{\partial
t} = &&\frac{3}{R}\frac{\partial}{\partial R}\left[\frac{R^{1/2}}{\Sigma}
\frac{\partial}{\partial R}
(\nu_1\Sigma R^{1/2}){\bf L}\right]\\
&+&\frac{1}{R}\frac{\partial}{\partial R}\left[\left(\nu_2 R^2\left| 
\frac{\partial{\bf l}}{\partial R}\right|^2-\frac{3}{2}\nu_1\right)
{\bf L}\right]\\
\nonumber&+&\frac{1}{R}\frac{\partial}{\partial
R}\left(\frac{1}{2}\nu_2R|{\bf L}|\frac{\partial{\bf l}}{\partial R}
\right).
\end{eqnarray}
In this equation, the terms proportional to $\nu_1$ describe the standard
viscous evolution of a thin and flat disc, where $\nu_1$ is the usual
viscosity acting on the $(R, \phi)$ shear. If we use the
$\alpha$-parametrization, we have
\begin{equation}
\nu_1=\alpha c_{\rm s}H = \alpha\Omega H^2.
\end{equation}

The terms proportional to $\nu_2$ in Equation (\ref{eq:pringle}) arise
whenever the disc is warped and $|\partial{\bf l}/\partial R|\neq 0$.
According to Equation (\ref{eq:pringle}) the warp diffuses with a diffusion
coefficient $\nu_2$. In view of this we shall define a second parameter
$\alpha_2$ defined similarly so that
\begin{equation}
\nu_2=\alpha_2 c_{\rm s}H = \alpha_2\Omega H^2.
\end{equation}
It is clear that the nature of the evolution of a warped accretion disc is
determined mainly by the relative values of $\alpha$ and $\alpha_2$.

We stress that Equation (\ref{eq:pringle}) was not derived from any
hydrodynamical (or magneto-hydrodynamical) equations. Rather the equation
results simply from the conservation equations (mass and angular momentum),
coupled with the assumption that the warp propagates in a diffusive manner.
Thus there is no linear approximation required in the derivation and the
equation is, in principle, applicable to warps of any amplitude, with an
appropriate choice of $\nu_2$.  Thus, equation (\ref{eq:pringle}) merely
states that the warp evolution is diffusive, with a diffusion coefficient
$\nu_2$ to be determined.

A number of authors have addressed the problem of warp diffusion.
\citet{pappringle83} considered the internal hydrodynamics of the disc,
assuming an internal {\it isotropic} viscosity for a warp small enough to be
treated as a linear perturbation (i.e. $\psi \ll H/R$), and for viscosity such
that $H/R \lesssim \alpha \ll 1$. In this case the internal hydrodynamic flow
is a laminar one, and is relatively straightforward to analyse (See Section
4). They found that to first order Equation (\ref{eq:pringle}) gives the
correct behaviour, along with the rather surprising result that
\begin{equation}
\frac{\nu_2}{\nu_1}=\frac{\alpha_2}{\alpha} = \frac{1}{2\alpha^2}.
\label{eq:alpha2}
\end{equation}
This implies that
\begin{equation}
\label{eq:prediction}
\alpha_2 = \frac{1}{2 \alpha},
\end{equation}
and therefore that the warp diffusion coefficient is {\it inversely}
proportional to the size of the viscosity. We discuss the physics behind this
in Section \ref{theory}.  To next order in $\alpha$ there are additional terms
which imply precession of the disc, and which are not included in the simple
formulation of Equation~(\ref{eq:pringle}).

\citet{ogilvie99} (see also \citealt{ogilvie00,ogilviedubus}) extends these
approximate analytic results by use of an asymptotic expansion in terms of the
small quantity $H/R$, but retaining the assumption of an isotropic
(Navier-Stokes) viscosity. By this means he is able to take account of larger
values of $\alpha$ and $\psi$ and to this approximation the internal dynamics
still in the form of well-ordered, laminar flows. \citet{ogilvie99} finds
\begin{equation}
\frac{\nu_2}{\nu_1}=\frac{1}{2\alpha^2}\frac{4(1+7\alpha^2)}{4+\alpha^2},
\end{equation}
when $\psi \ll 1$ and also gives higher order corrections with terms
in $\psi^2$. He also finds additional precessional terms, which are
generally of higher order than the terms already included (We discuss
this further in Section \ref{sec:prec}).

None of these studies takes into account the fact that viscosity in real
accretion discs is most likely not a simple isotropic Navier-Stokes viscosity.
It is generally believed (see for example \citealt{balbusreview}) that the
mechanism responsible for angular momentum transport in discs (corresponding
to $\nu_1$) is MHD turbulence, driven by the magneto-rotational instability.
An attempt at estimating $\nu_2$ has been made by \citet{torkelsson00} in a
numerical shearing box simulation. It has become evident recently, however,
that the results of such simulations are somewhat problematic
\citep{king07,pessah07,fromang07}. In addition, as discussed by Pringle
(1992), there is no particular reason to assume that the mechanism which damps
the warp (corresponding to $\nu_2$) is the same.  In addition, for large
(astrophysically significant) warps the internal motions can become
sufficiently large, with relative supersonic velocities, that a number of
authors have argued that they are likely to be unstable
\citep{coleman92,gammie00} and so result in enhanced dissipation.

In the opposite regime of thick discs, or of lower viscosity, the warp evolves
to first order as a non-dispersive bending wave \citep{paplin95}. The
equations of motion for a wave in the case where the disc is Keplerian and
nearly inviscid are \citep{lubow00,lubow02}

\begin{equation}
\Sigma R^3\Omega \frac{\partial{\bf l}}{\partial t} = \frac{\partial{\bf G}}
{\partial R}
\label{eq:wave1}
\end{equation}
\begin{equation}
\frac{\partial{\bf G}}{\partial t} + \alpha\Omega{\bf G} = 
\Sigma R^3\Omega \frac{c_{\rm s}^2}{4}\frac{\partial{\bf l}}{\partial R},
\label{eq:wave2}
\end{equation}
where ${\bf G}$ is the disc internal torque. The corresponding dispersion
relation is then \citep{nelson99}
\begin{equation}
\omega(\omega-i\alpha\Omega)=\frac{c_{\rm s}^2k^2}{4},
\end{equation}
where $\omega$ is the wave frequency, $k$ the wavenumber and $c_{\rm
  s}$ the sound speed. The solution for $\omega$ is
\begin{equation}
\omega = \frac{1}{2} \{ i \alpha \Omega \pm [ c_{\rm s}^2 k^2 -
  \alpha^2 \Omega^2]^{1/2} \}.
\end{equation}
Thus the above relation shows that in the absence of viscosity
($\alpha = 0$) a bending wave propagates at half the sound speed. When
$\alpha \ge H k$, i.e. when $\alpha \ge 2 \pi H /\lambda$, where
$\lambda$ is the wavelength of the perturbation, we see that $\omega$
is purely imaginary, and wave-like propagation no longer
occurs. Propagation becomes purely diffusive in the limit $|\omega| \ll
\alpha \Omega$, in which case we obtain
\begin{equation}
\frac{\omega}{i\alpha\Omega}=\left(\frac{c_{\rm
      s}k}{2\alpha\Omega}\right)^2\ll 1,
\end{equation}
Thus pure diffusion occurs when $\alpha \gg \frac{1}{2} H k = \pi H /
\lambda$. The boundary between the purely diffusive and purely wave-like
propagation regimes is not a sharp one. For convenience we shall take the
transition between diffusive and wave-like regimes to occur at the value
$\alpha = \alpha_{\rm c}$ when the real and imaginary parts of $\omega$ are
equal $\omega_{\rm R} = \omega_{\rm I}$. Thus the critical value of $\alpha$
is given by
\begin{equation}
\alpha_{\rm c} = \frac{1}{\sqrt{2}}Hk.
\end{equation}

In this paper we investigate the hydrodynamics of diffusive warp propagation
(i.e. $\alpha > \alpha_{\rm c}$) using numerical simulations. In particular,
we use Smoothed Particle Hydrodynamics (SPH). Such a technique has been used
in some of the most accurate investigations of warp propagation to date
\citep{nelson99,nelson00,larwood96} and is particularly suited to the
treatment of complex and variable geometries, like that of a warped disc.
However, as we will further discuss below, some numerical effects, for example
related to the implementation of viscosity, might play a role and we stress
the importance of conducting similar analyses with different codes, especially
in view of simulating a warped disc that evolves under the action of MHD
instabilities, which are not taken into account here. The early investigations
by \citet{nelson99,nelson00} have focussed on the thick disc regime, where the
warp propagates as a bending wave, but have already noticed some dissipational
effects. In this respect, the present work can be seen as complementary to
these previous ones, since we here focus on the thin disc regime, where the
character of warp propagation is expected to differ substantially from the
thick disc case.

The paper is organized as follows. In Section 2 we present the numerical
setup, and we present the results in Section 3. In Section 4 we present a
simplified discussion of the physics of diffusive warp propagation, and use
this analysis to interpret our results. We discuss the implications and
limitations in Section 5.

\begin{table*}
  \centering
  \begin{tabular}{ccccccccccc}
    Sim  & $H/R$ & No gas particles 
    & $\psi_{\rm max}$ & $\alpha$ & $f$ & $\alpha_2$ & $\alpha_3$ & $\langle h\rangle/H$ 
    & VS & $\alpha_{\rm SPH}$\\
    \hline
    S0 & 0.0133 & $2~10^6$ & 0.026 & 0.28 & 1    &1.78 &-0.51 & 0.6 & on & 15\\
    S0b& 0.0133 & $2~10^6$ & 0.026 & 0.28 & 1    &1.78 & 0.08 & 0.6 & off &7.5\\
    S1 & 0.0133 & $2~10^6$ & 0.026 & 0.23 & 1    &2.17 &-0.46 & 0.6 & on  &10\\
    S1b& 0.0133 & $2~10^6$ & 0.026 & 0.23 & 1    &2.17 & 0.17 & 0.6 & off &5\\
    S2 & 0.0133 & $2~10^6$ & 0.026 & 0.18 & 1    &2.77 &-0.72 & 0.6 & on &7.5 \\
    S3 & 0.0133 & $2~10^6$ & 0.026 & 0.15 & 0.85 &2.83 &-0.6  & 0.6 & on &5 \\
    S4 & 0.0133 & $2~10^6$ & 0.026 & 0.14 & 0.8  &2.85 & 0.22 & 0.6 & off &2.5\\
    S5 & 0.0133 & $2~10^6$ & 0.026 & 0.09 & 0.6  &3.33 &-0.63 & 0.6 & on &2.5 \\
    S6 & 0.0133 & $2~10^6$ & 0.026 & 0.07 & 0.42 &3    & 0.21 & 0.6 & off&1.7\\
    S7 & 0.0133 & $2~10^6$ & 0.026 & 0.05 & 0.38 &3.8  & 0.2  & 0.6 & off&1 \\
    S8 & 0.0133 & $2~10^6$ &  1.3  & 0.26 & 0.75 &1.44 & 0.52 & 0.6 & off&5 \\
    S9 & 0.0133 & $2~10^6$ &  1.3  & 0.1  & 0.6  &3    &-0.17 & 0.6 & on &2.5 \\
    S10& 0.0334 &  $10^6$  & 0.026 & 0.1  & --   &--   & --   & 0.4 & on &10\\
    S11& 0.0334 &  $10^6$  & 0.026 & 0.08 & --   &--   & --   & 0.4 & on &7.5\\
    S12& 0.0334 &  $10^6$  & 0.026 & 0.06 & --   &--   & --   & 0.4 & on &5\\
    S13& 0.0334 &  $10^6$  & 1.3   & 0.08 & --   &--   & --   & 0.4 & on &7.5\\
    S14& 0.0668 &  $10^6$  & 1.3   & 0.04 & --   &--   & --   & 0.1 & on &5\\
    \hline
  \end{tabular}
  \caption{\small Summary of the main physical properties of the simulations
  carried out. Here $H/R$ refers to the value calculated at $R=R_0$. See text
  (section \ref{sec:results}) for the definition of the parameter $f$ and
  for details of the calibration of the disc viscosity parameter
  $\alpha$. Note that $\alpha_2 = f/2 \alpha$.}
  \label{tab:table}
\end{table*}
\begin{figure}
\centerline{\epsfig{figure=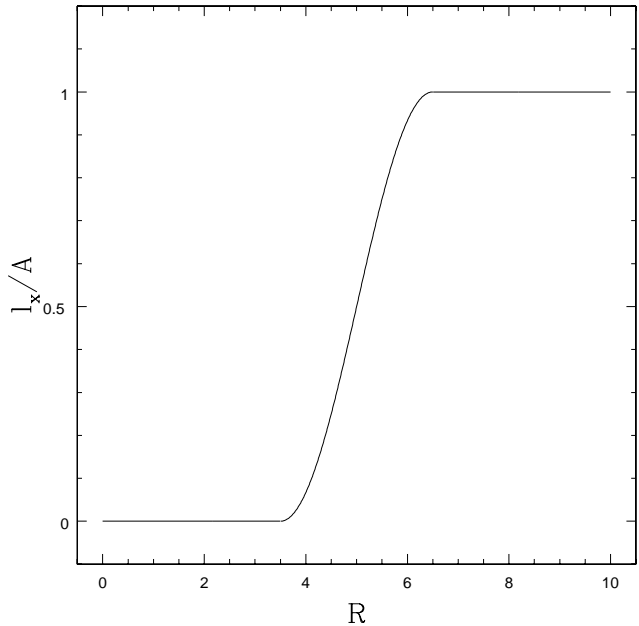,width=0.5\textwidth}}
\caption{Initial profile of the $x$-component of the unit vector {\bf l} in
  our SPH simulations. Note that the $y$-component is initially zero,
  so that the initial warp has no twist.}
\label{fig:initial}
\end{figure}

\begin{figure*}
\centerline{\epsfig{figure=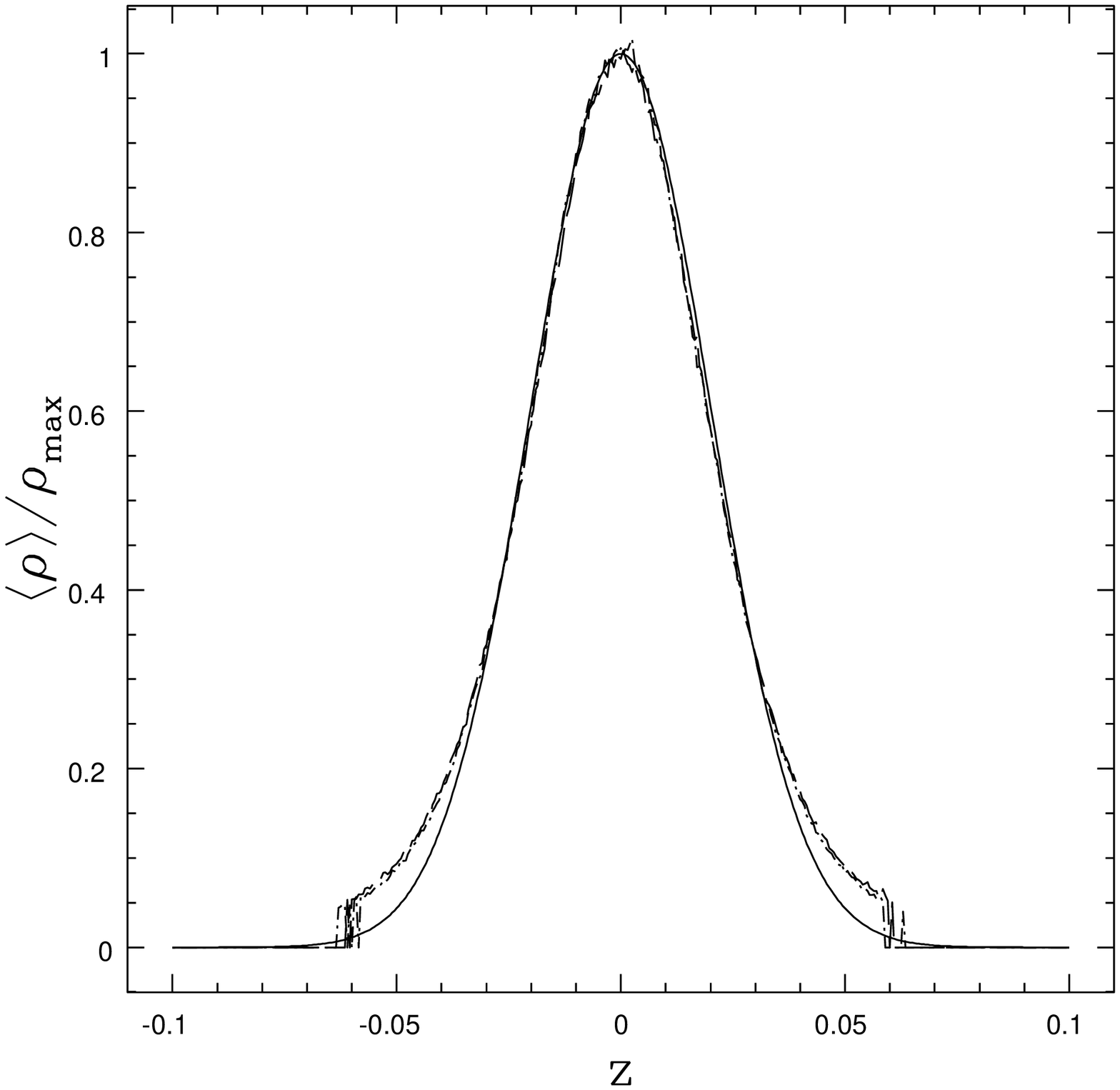,width=0.5\textwidth}
            \epsfig{figure=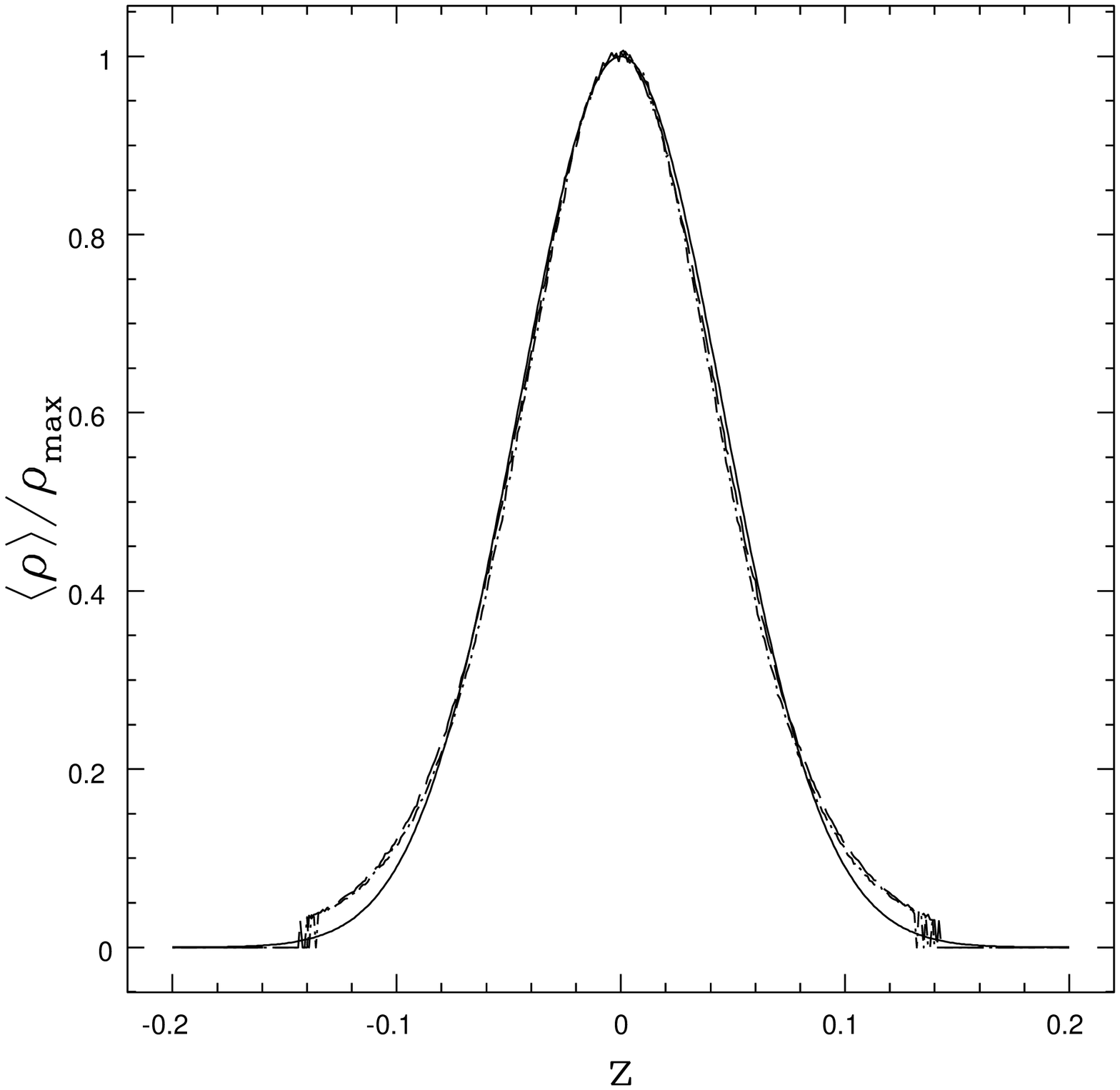,width=0.5\textwidth}}
          \caption{Azimuthally averaged vertical density profile for
            simulations S0 to S9, at $R=1$, where the disc thickness $H=0.02$
            (left) and $R=3$, where $H=0.045$ (right).  The solid line shows
            the theoretical Gaussian profile expected in vertical hydrostatic
            balance, while the dashed and dotted lines (almost exactly
            coincident) show the density evaluated from the SPH code at $t=10$
            and $t=30$ code units, respectively. Deviations from the expected
            profile only occur at a height $z\gtrsim 2H$, where the density
            has decreased by a factor 0.2.}
\label{fig:dens}
\end{figure*}

\section{Numerical setup}

We have performed a set of three-dimensional SPH simulations of warped discs.
SPH is a Lagrangian hydrodynamic scheme
\citep{lucy77,gingold77,benz90,monaghan92}, in which the gas disc is modeled
with $N$ particles. For most of our simulations we have used 1 or 2 million
particles (these are some of the highest resolution SPH disc simulations to
date).

In the following we use code units in which the gravitational constant
$G=1$, the central point mass $M=1$, so that at a radius $R=1$ in code
units, the dynamical time is $\Omega^{-1} = 1$.

We set up our warped disc by placing the gas particles in Keplerian orbits
around a central point mass $M=1$ (in code units), modeled as a sink particle
\citep{bate95}, with accretion radius $R=0.5$ in code units. We distribute
them in such a way that the disc has a prescribed initial surface density
profile, as described below. We assign each particle a radius dependent sound
speed $c_{\rm s}$, so as to attain the desired temperature profile (see
below). The particles are then distributed in $z$ so as to attain a Gaussian
density profile in the vertical direction, with thickness $H=c_{\rm
  s}/\Omega$. We then tilt the orbit of each particle in such a way that the
components of the unit vector ${\bf l}$ are given by
\begin{equation}
l_{x}= \left\{ \begin{array}{ll}
    \displaystyle 0  & \hspace{0.1cm}
\mbox{for} \hspace{0.1cm} R<R_1 \\
\\
    \displaystyle
    \frac{A}{2}\left[1+\sin\left(\pi\frac{R-R_0}{R_2-R_1}\right)\right]
      & \hspace{0.1cm}
\mbox{for} \hspace{0.1cm} R_1<R<R_2\\
\\
    \displaystyle A  & \hspace{0.1cm}
\mbox{for} \hspace{0.1cm} R>R_2 
\end{array}\right.
\end{equation}
\begin{equation}
l_{\bf y} = 0,
\label{eq:initial}
\end{equation}
\begin{equation}
l_{z} = \sqrt{1-l_{x}^2},
\end{equation}
where $R_1=3.5$ and $R_2=6.5$ in code units, and $R_0 = (R_1 +R_2)/2$.  The
initial shape of the warp is plotted in Fig. \ref{fig:initial}. The warp
amplitude $\psi$ is then
\begin{equation}
  \psi=R\left|\frac{\partial{\bf l}}{\partial R}\right|=\frac{R}{l_{z}}
\frac{\partial l_{x}}{\partial R}
\end{equation}
the maximum of which is attained at $R\approx R_0$ and is given by
\begin{equation}
\psi_{\rm max}\approx \frac{\pi R_0 A}{2(R_1-R_2)}=2.62 A
\end{equation}

The disc extends from $R_{\rm in}=0.5$ to $R_{\rm out}=10$, with a surface
density profile, $\Sigma$, given by
\begin{equation}
\Sigma(R)=\Sigma_0 R^{-p}\left(1-\sqrt{\frac{R_0}{R}}\right).
\end{equation}
Note that the density normalization $\Sigma_0$ does not play a role in our
simulations, since we do not include the disc self-gravity in the computation.
The parameter $p$ is set to $p=3/2$. The simulations we perform are locally
isothermal, with a sound speed $c_{\rm s}$, given by
\begin{equation}
c_{\rm s}=c_{\rm s,0} R^{-q},
\end{equation}
where $q=3/4$ and where the normalization $c_{\rm s,0}$ determines the disc
thickness. 

In order to model the viscous evolution of the disc, we use the SPH artificial
viscosity formalism \citep{monaghan92}. It has been shown
\citep{lubow94,murray96} that artificial viscosity in SPH can mimic the
behaviour of an $\alpha$-viscosity, with
\begin{equation}
\alpha \propto \alpha_{\rm SPH} \frac{\langle h\rangle}{H},
\end{equation}
where $\alpha_{\rm SPH}$ is the artificial viscosity coefficient, $\langle
h\rangle$ is the average smoothing length and $H$ is the disc thickness. Our
setup has been chosen in such a way that the disc thickness is uniformly
resolved at different radii. Indeed, the disc thickness $H$ scales with radius
as
\begin{equation}
H=\frac{c_{\rm s}}{\Omega}\propto R^{3/2-q},
\end{equation}
and we expect the average smoothing length at one radius to scale as
\begin{equation}
\langle h\rangle \propto \rho^{-1/3} \propto
\left(\frac{\Sigma}{H}\right)^{-1/3} \propto R^{(p-q)/3+1/2}.
\end{equation}
Thus, by choosing $p=3/2$ and $q=3/4$, we have that $H\propto \langle h\rangle
\propto R^{3/4}$. This then produces a disc with a constant $\alpha$, the
magnitude of which can be adjusted by varying the SPH artificial viscosity
parameter $\alpha_{\rm SPH}$. Additionally, since $\nu_1=\alpha\Omega H^2$,
and $H\propto R^{3/4}$, we also expect that the simulations be characterized
by a constant viscosity $\nu_1$.\footnote{We thank Gordon Ogilvie for
  suggesting this setup to us.} As for the quadratic term in the standard
formulation of SPH artificial viscosity (the `$\beta$'-term), we have only
included it in simulations for which the `viscous switch' is on (see below),
in which cases we have adopted the common practice of setting $\beta_{\rm
  SPH}=2\alpha_{\rm SPH}$.

We have performed several simulations, varying the disc aspect ratio
$H/R$, the viscosity $\alpha$ (through varying the SPH artificial
viscosity $\alpha_{\rm SPH}$, see Section \ref{sec:results} below for
details on how we calibrate the viscosity) and the peak warp amplitude
$\psi$. The main parameters of our simulations are summarized in Table
\ref{tab:table}. The ninth column of Table \ref{tab:table} shows the
(radius-independent) ratio of the azimuthal and vertical average of
the smoothing length $h$ and the disc thickness $H$, which also
demonstrates that we resolve the vertical structure moderately well in
the simulations. Note that this column shows a vertical average of the
smoothing length $h$, including the contribution of particles at high
$z$ which have a smaller density and hence a larger $h$. The actual
value of $h$ in the bulk of the disc can be significantly smaller than
the average. As a further check of the resolution achieved in our
simulations, we plot in Fig.  \ref{fig:dens} the azimuthally averaged
vertical density profile for simulations S0 to S9 (see Table
\ref{tab:table}), at two different radii, $R=1$ (left panel) and $R=3$
(right panel). The solid line shows the theoretical profile derived
applying the standard hydrostatic balance condition, while the dashed
and dotted lines (almost exactly coincident) show the value of the
density as computed from the SPH scheme at two times, $t=10$ (dashed)
and $t=30$ (dotted) code units. We can clearly see that the SPH
estimates follow nicely the predicted profile, deviating from it only
at vertical height $z\gtrsim 2H$, where the density is already a
factor $0.2$ smaller than the peak.

As can be seen from Table \ref{tab:table} we have considered three different
values for the disc thickness. Most of our simulations have $H/R\approx
0.0133$, a few have $H/R\approx 0.0334$ and only one had a relatively large
thickness of $H/R\approx 0.0668$. The critical value of $\alpha$ below which
wave propagation is expected to occur is then $\alpha_{\rm c}=0.05, 0.12$, and
0.25, respectively. It can be then easily seen that the thickest case we
consider falls in the wave propagation regime, while the thinnest case (the
majority of simulations) falls in the diffusion regime. The intermediate
thickness is marginal, and wave propagation with significant dissipation is
expected.

\begin{figure*}
\centerline{\epsfig{figure=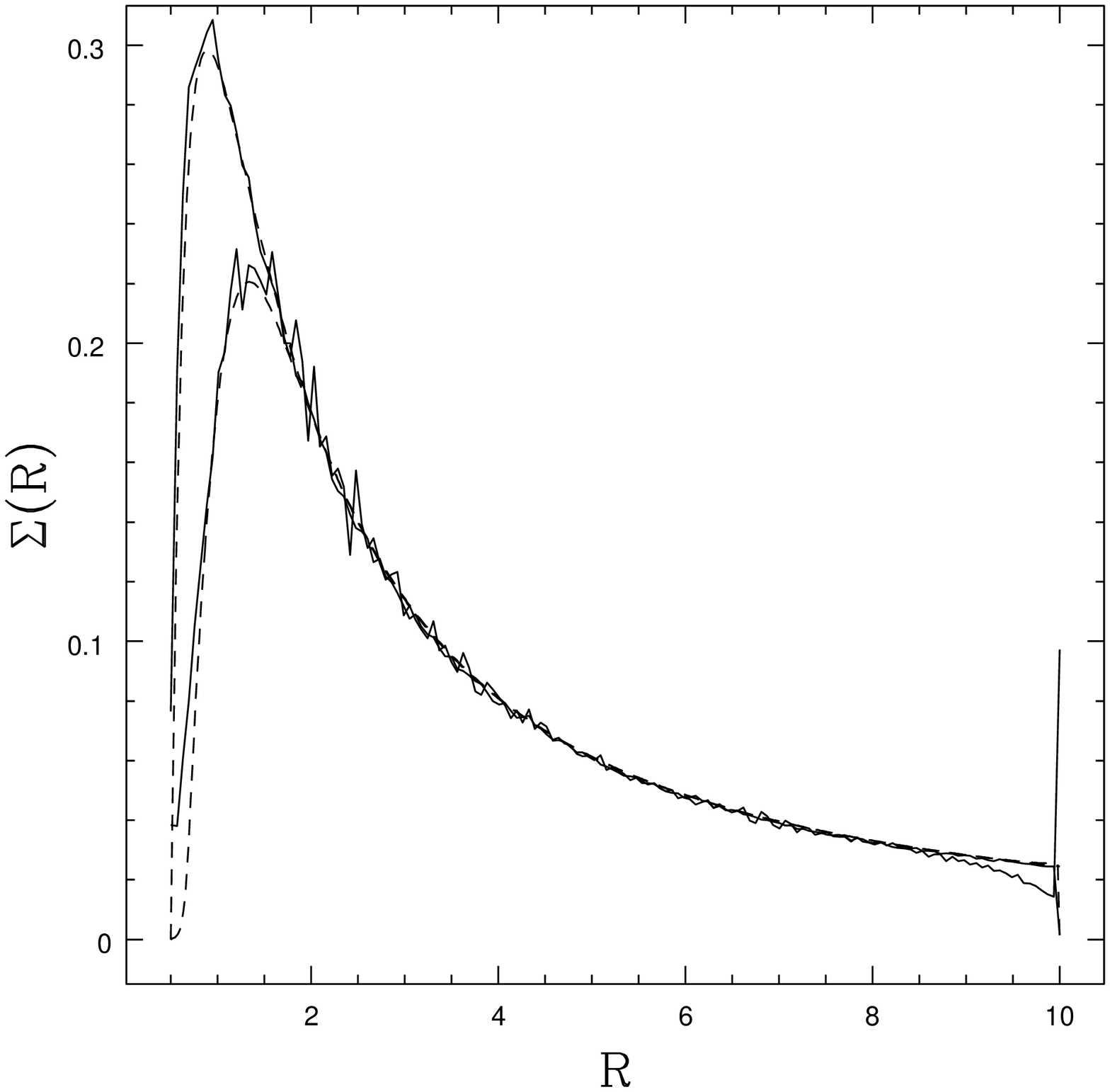,width=0.5\textwidth}
            \epsfig{figure=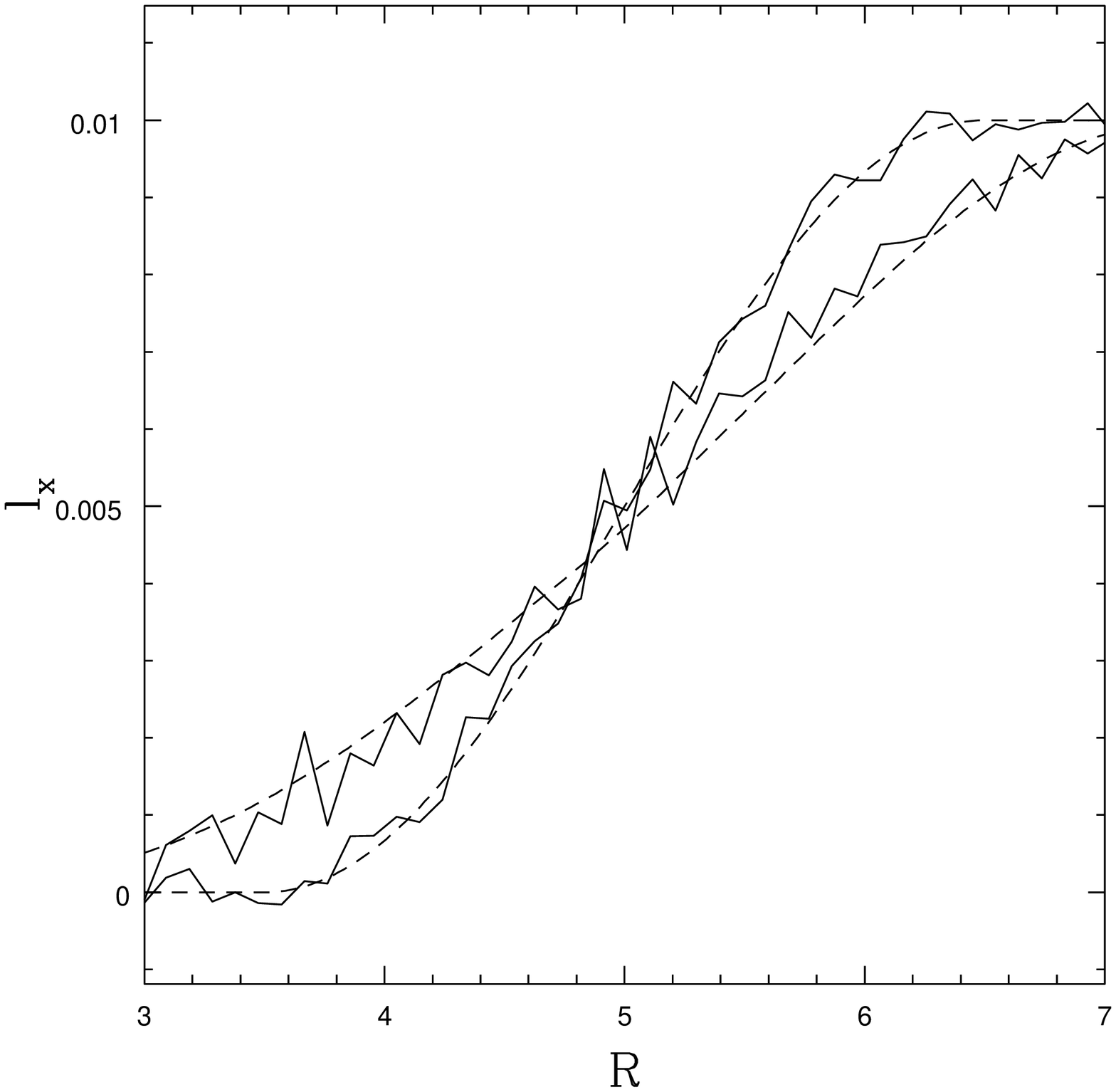,width=0.5\textwidth}}
          \caption{Evolution of simulation S11. The solid lines refer to the
            azimuthal and vertical average of the SPH simulation, while the
            dashed lines show the evolution of the corresponding initial
            conditions obtained by applying Eqs. (\ref{eq:wave1}) and
            (\ref{eq:wave2}) for the warp, while the surface density is
            evolved using a viscosity parameter $\alpha=0.08$. The left panel
            shows the evolution of the surface density $\Sigma$, while the
            right panel shows the evolution of $l_{x}$. The different
            lines refer to $t=0$ and to $t=155$ (in units of the dynamical
            time at $R=1$).}
\label{fig:S11}
\end{figure*}

\subsection{On modeling disc viscosity in SPH}

In this paper, we have used the internal SPH viscosity in order to
simulate a (Navier-Stokes) viscosity in the disc. This is clearly an
idealization, since there is no a priori expectation that real disc
viscosity in any way resembles Navier-Stokes. In practice accretion
discs transport are likely to be be dominated by other forms of stress
(for example, due to the magneto-rotational instability). But at least
this provides some insight on the evolution of the warp in this
specific case, and provides some comparison with previous analytic
work.

However, care should be taken when using the SPH artificial viscosity
to model a Navier-Stokes viscosity. Standard implementations of SPH
adopt the so-called `viscous switch', so that viscous forces are only
active when acting on approaching particles, and vanish for receding
particles. Clearly, if we want to simulate something approximating a
Navier-Stokes viscosity we should turn the `viscous switch' off.
\citet{lubow94} and \citet{murray96} have already remarked on this.

Because it is unclear how the real, physical disc viscosity behaves, we have
run simulations with and without the `viscous switch'. The column `VS' is
Table \ref{tab:table} indicates the simulations that had the switch on or off
(note that obviously in order to reproduce the same physical viscosity
coefficient $\alpha$, a much smaller value of $\alpha_{\rm SPH}$ -- roughly
half -- is needed when the viscous switch is off, as shown in Table
\ref{tab:table}.). This enables us to directly check the impact of using the
viscous switch or not in the evolution of our warped discs. As we discuss
below (Section 3.3), the effects on the evaluation of the warp diffusion
coefficient (i.e. on the diffusion/damping of the warp) are minor, if not
negligible. There is however, a strong sensitivity in the rate, and indeed
direction, of internal precession induced by the warp.

\section{Results}
\label{sec:results}

In this Section we describe the main results of our simulations.  Our
general procedure for the analysis is relatively simple. We first
divide the radial range of the simulation in a series of annuli at
different radii $R$. From the output of the SPH simulation we then
compute the surface density $\Sigma(R)$ and the azimuthal and vertical
average of the three components of the unit vector ${\bf l}$ as
functions of time. We then compare the evolution of these quantities
with the evolution predicted either by Eq. (\ref{eq:pringle}) (for the
thinnest cases), or by Eqs. (\ref{eq:wave1}) and (\ref{eq:wave2}) (for
the thick cases).

\subsection{Wave-like propagation in the thick disc case}

As mentioned above, we expect the warp to propagate as a bending wave in the
thickest of our simulations, and possibly also the intermediate thickness
ones. This is indeed confirmed by our simulations. As an example, we show in
Fig. \ref{fig:S11} the evolution of simulation S11. The left hand panel shows
the evolution of the surface density $\Sigma$ (solid lines), compared to the
evolution predicted by a simple viscous surface density evolution with
$\alpha=0.08$.  The right hand panel shows the evolution of the $x$-component
of ${\bf l}$ (solid lines) compared to the evolution predicted from Eqs.
(\ref{eq:wave1}) and (\ref{eq:wave2}). Note that in order to obtain a good fit
to the simulations it is essential also to include the dissipative term in Eq.
(\ref{eq:wave2}), i.e. the second term on the left-hand side of the equation,
since a pure wave evolution does not fit the data. We also note that the same
evolution could also in principle be fitted (at least over the limited time
span covered by the simulation) with a diffusive warp propagation model, but
where the warp diffusion coefficient is taken to be much smaller than the
value predicted from the linear theory, i.e.  $\alpha_2\approx 2.5\ll
1/2\alpha$.

\begin{figure*}
\centerline{\epsfig{figure=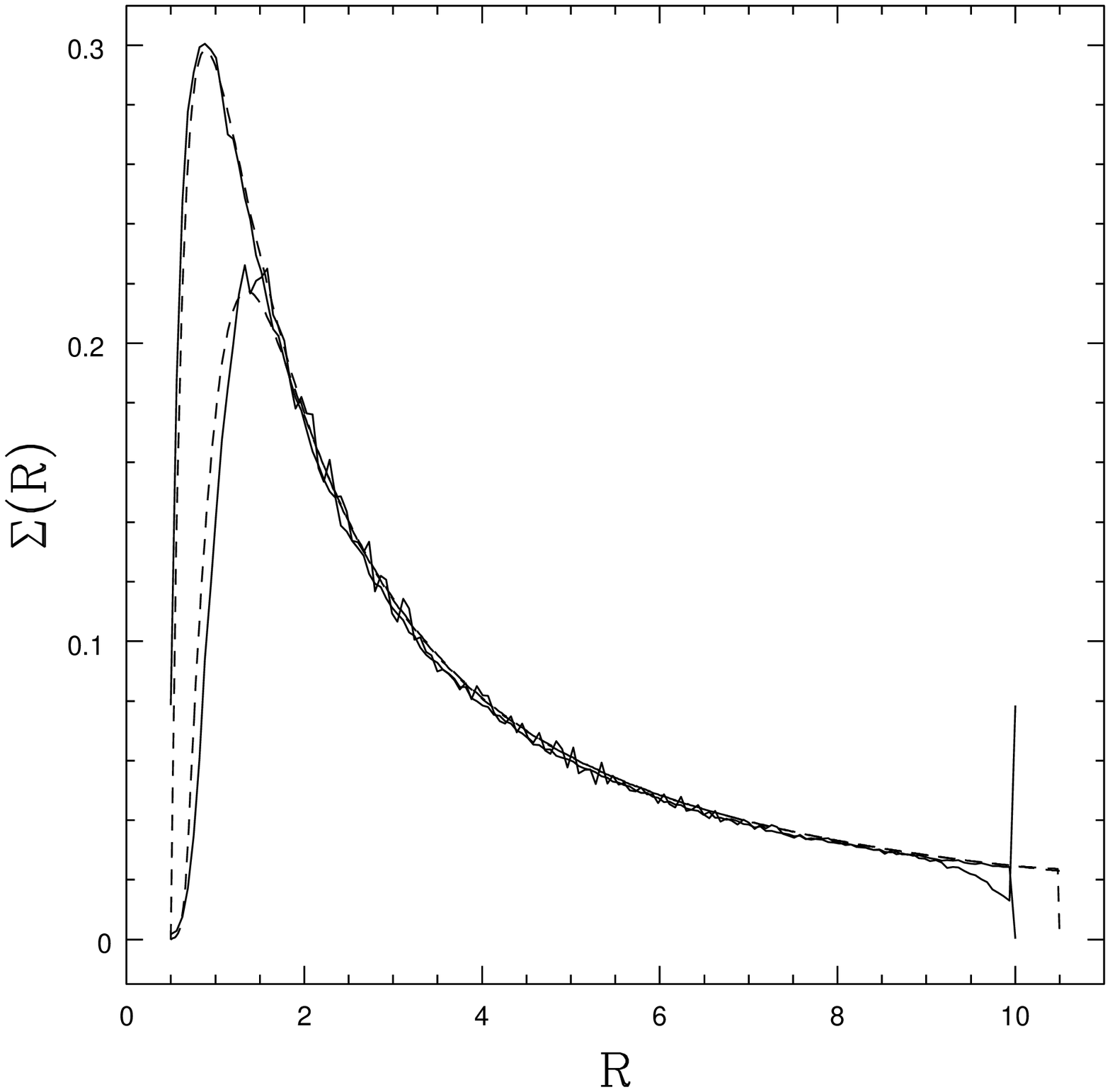,width=0.5\textwidth}
            \epsfig{figure=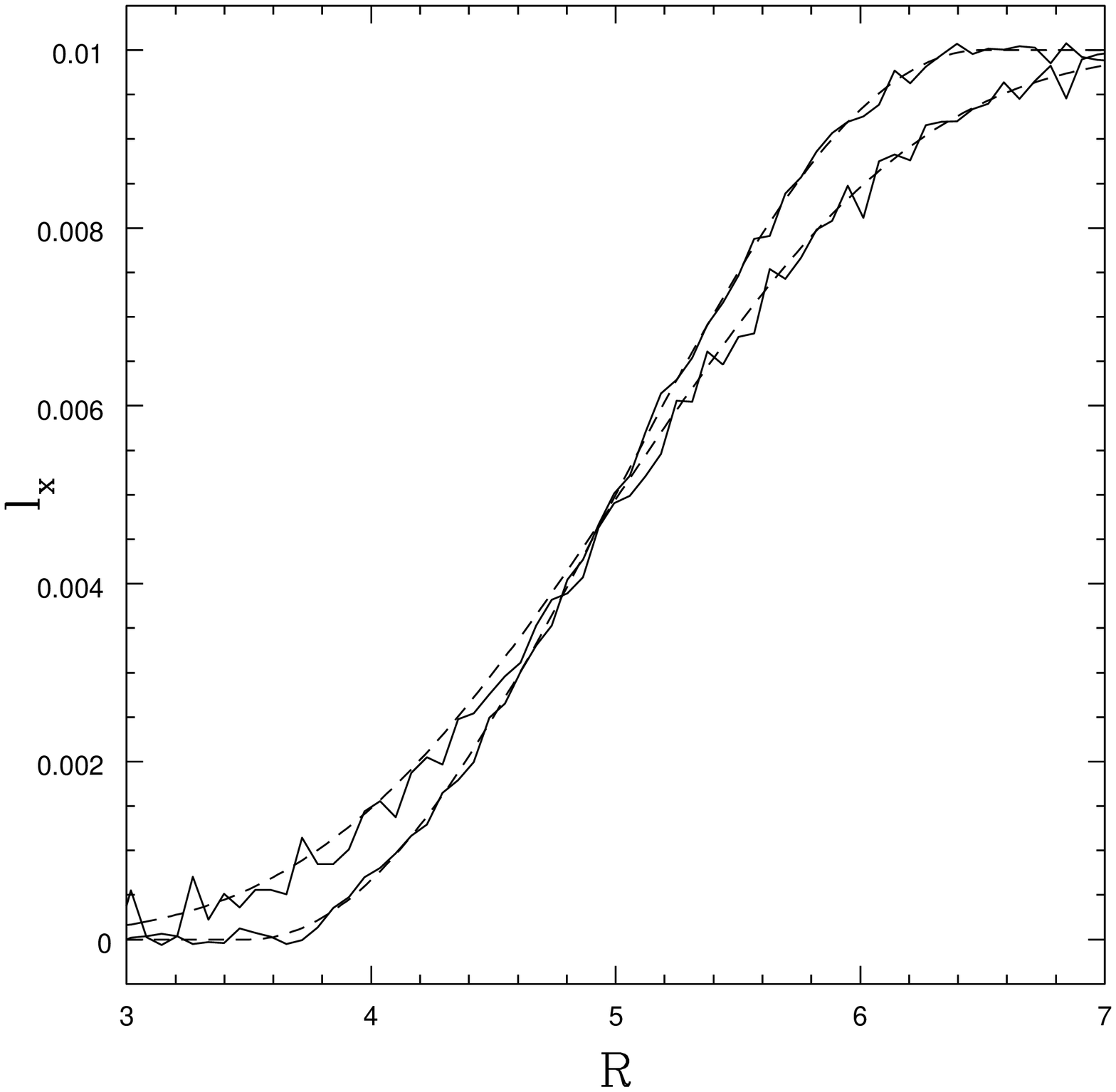,width=0.5\textwidth}}
\caption{Evolution of simulation S2. The solid lines refer to the azimuthal
  and vertical average of the SPH simulation, while the dashed lines show the
  evolution of the corresponding initial conditions obtained by applying
  Eq. (\ref{eq:pringle}), with the following parameters: $\alpha=0.18$ and
  $f=1$. The left panel shows the evolution of the surface density $\Sigma$,
  while the right panel shows the evolution of $l_{x}$. The different
  lines refer to $t=0$ and to $t=465$ (in units of the dynamical time at
  $R=1$).}
\label{fig:S2}
\end{figure*}

\begin{figure*}
\centerline{\epsfig{figure=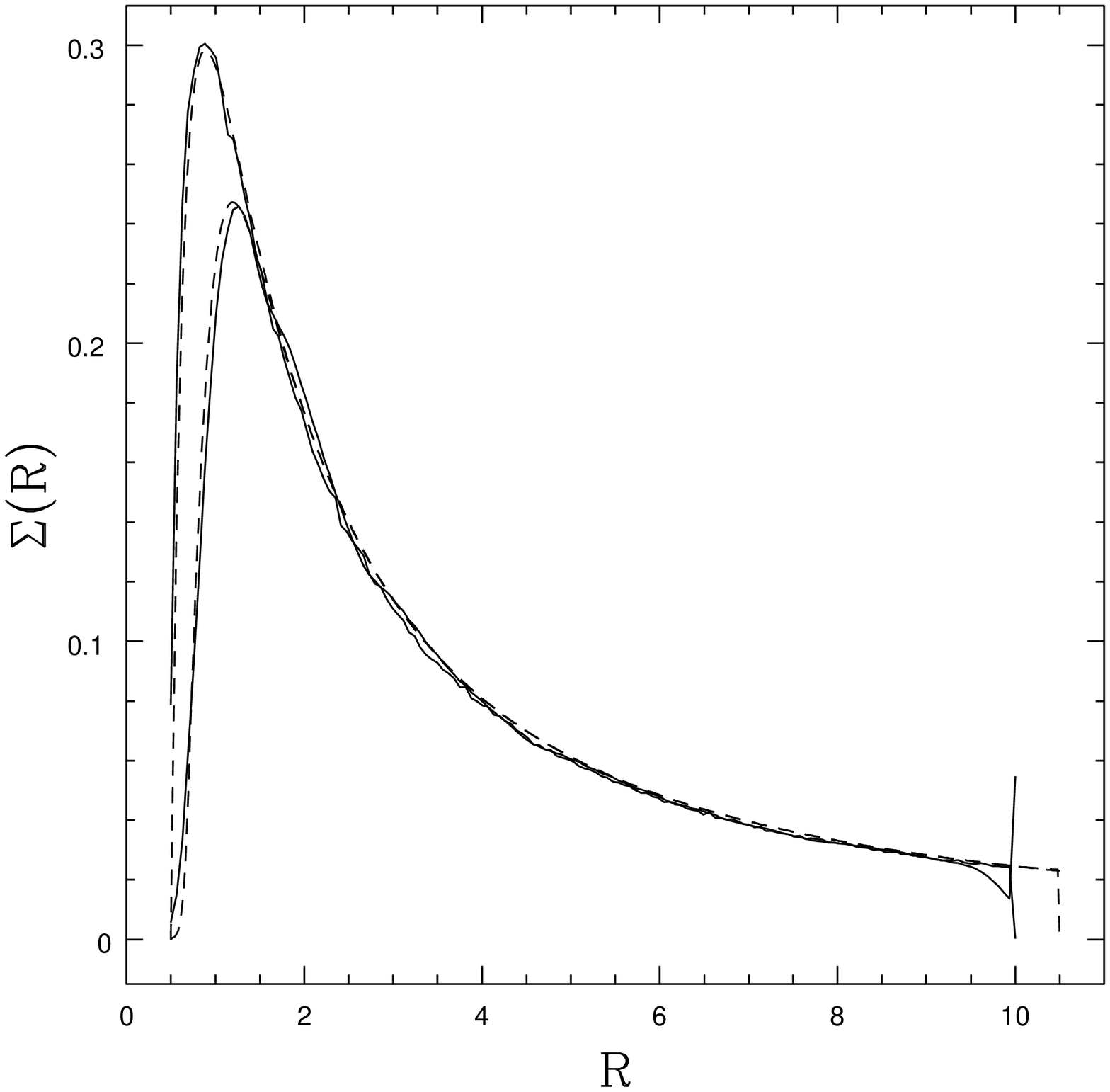,width=0.5\textwidth}
            \epsfig{figure=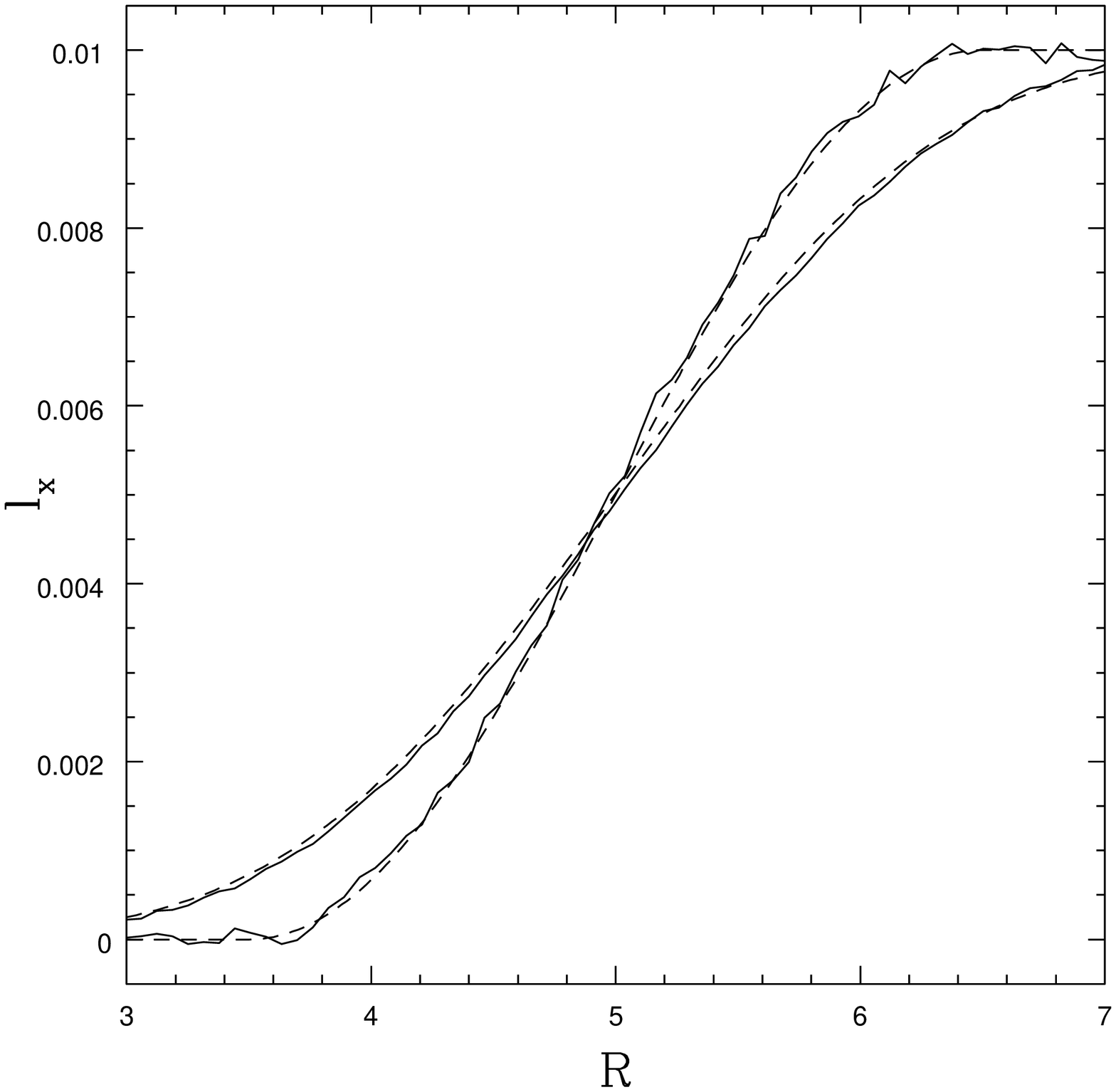,width=0.5\textwidth}}
          \caption{Evolution of simulation S6. The solid lines refer to the
            azimuthal and vertical average of the SPH simulation, while the
            dashed lines show the evolution of the corresponding initial
            conditions obtained by applying Eq. (\ref{eq:pringle}), with the
            following parameters: $\alpha=0.07$ and $f=0.42$. The left panel
            shows the evolution of the surface density $\Sigma$, while the
            right panel shows the evolution of $l_{x}$. The different
            lines refer to $t=0$ and to $t=550$ (in units of the dynamical
            time at $R=1$).}
\label{fig:S6}
\end{figure*}

\subsection{Diffusive propagation in the thin disc case}

For our thinnest disc case we expect the warp to propagate diffusively,
according to Eq. (\ref{eq:pringle}). We have then solved Eq.
(\ref{eq:pringle}) numerically using the techniques described in detail in
\citet{pringle92} and in \citet{LP06}. The model described in equation
(\ref{eq:pringle}) contains two free parameters, namely, the two viscosities
$\nu_1$ and $\nu_2$, and we adjust these in order to match the evolution of
the simulation. In particular, the surface density $\Sigma$ is most sensitive
to the value of the viscosity $\nu_1$, while the evolution of the component
$l_{x}$ of the unit vector ${\bf l}$ is most sensitive to the value of
$\nu_2$. We therefore use these two quantities separately to determine the two
parameters. To be sure, the evolution of $\Sigma$ does also depend on $\nu_2$
\citep{pringle92}. However, this effect scales quadratically with the warp
amplitude and can be generally neglected in most of our simulations (with some
exceptions, as shown below).

In practice, we determine the value of $\alpha$ and the value of a
parameter $f$, which is a measure of the deviation of $\nu_2$ from the value
expected from the linear theory. Thus $f$ is defined by

\begin{equation}
\frac{\nu_2}{\nu_1}=\frac{f}{2\alpha^2}.
\end{equation}

The fifth and sixth columns of Table \ref{tab:table} show the measured values
of $\alpha$ and $f$ for all of our simulations. It can be seen that we are
able to span a range of $\alpha$ that goes from $0.05$ to $0.28$, that is
roughly a factor 7. Only for the three largest values of $\alpha$ does the
measured value of $f$ agree with the linear estimate, while for most other
cases the diffusion coefficient $\nu_2$ is smaller than predicted by
linear theory.

Note that, with regard to the evaluation of the warp diffusion
coefficient, the simulations which have an implementation of viscosity
that more closely resembles a Navier-Stokes viscosity (i.e. viscous
switch off) do not show significant differences with respect to the
corresponding ones with the viscous switch turned on. Simulations S0
and S0b and S1 and S1b, that have the same estimated value of
$\alpha$, also share the same warp diffusion coefficient. Simulations
S3 and S4, which have very similar values of $\alpha$, also have a
similar diffusion coefficient. Note that this agreement encompasses
the whole range of viscosities that we probe in the thinnest disc
configuration, extending from high values of $\alpha$, for which the
warp diffusion rate agrees with the linear theory, down to lower
values, where some non-linear effects start to play a role.

\begin{figure}
\centerline{\epsfig{figure=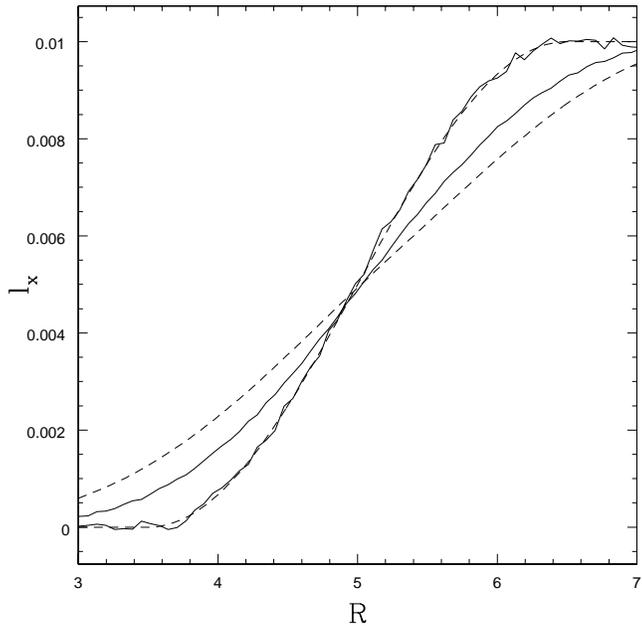,width=0.5\textwidth}}
\caption{Evolution of simulation S6. The solid lines refer to the azimuthal
  and vertical average of the SPH simulation, while the dashed lines show the
  evolution of the corresponding initial conditions obtained by applying Eqs.
  (\ref{eq:wave1}) and (\ref{eq:wave2}), with $\alpha=0.07$. The different
  lines refer to $t=0$ and to $t=550$ (in units of the dynamical time at
  $R=1$).}
\label{fig:S6wave}
\end{figure}

\begin{figure}
\centerline{\epsfig{figure=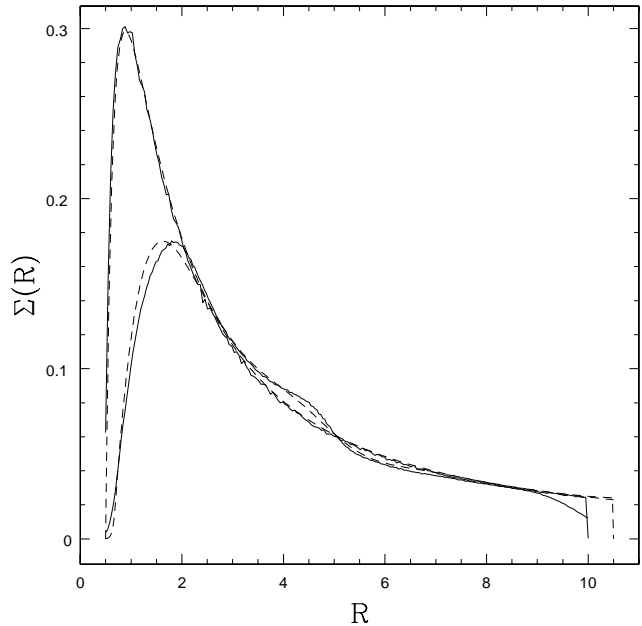,width=0.5\textwidth}}
\caption{Evolution of simulation S8. The solid line shows the results of the
  SPH simulations, while the dashed line shows the diffusive evolution,
  according to Eq. (\ref{eq:pringle}). The two snapshots refer to $t=0$ and to
  $t=870$ in code units. The inflection in the surface density is caused by
  the effect of $\nu_2$ on the surface density, as discussed by
  \citet{pringle92}.}
\label{fig:S8}
\end{figure}

Fig. \ref{fig:S2} shows one example of our analysis, corresponding to
the case where $H/R=0.0133$, $\alpha=0.18$ and $\psi_{\rm max}=0.026$
(simulation S2).  The left panel shows the evolution of $\Sigma$,
while the right one shows the evolution of $l_{x}$. The solid
lines refer to the results of the SPH evolution at two different
times, $t=0$ and $t=465$ in code units (where the unit time is the
dynamical time at $R=1$), while the dashed lines show the evolution of
the simple model of Equation (\ref{eq:pringle}). As can be seen, the
evolution of the two quantities is well reproduced by the model,
therefore demonstrating numerically for the first time the validity of
the diffusive model for warp propagation in this regime. In this
particular case, not only does the warp evolve diffusively, but the
value of the diffusion coefficient agrees with the expectations from
the linear theory, that is, we measure $f=1$.

Fig. \ref{fig:S6} shows the evolution of simulation S6.  This
simulation is also well fitted by a diffusive model, except that in
this case, although the value of $\alpha = 0.07$ is above the critical
value so that diffusive propagation is expected (in this case
$\alpha_{\rm c}\simeq 0.05$), the required value of $\alpha_2\approx
3$ is significantly below the value expected from the linear theory
$1/2\alpha\approx 7.14$. To be sure that wave propagation effects do
not play a significant role here, we have also tried to fit the
evolution of this simulation with a wave propagation model, as done in
the previous section. The results are shown in
Fig. \ref{fig:S6wave}. It can be seen that in this case a wave
propagation model (including the effects of viscosity) does not
reproduce the results of the simulation.

Most of our simulations have a low amplitude warp, with $\psi\approx
0.026$.  However, we have also run some simulations with $\psi\sim
1$. In the thin disc case, these simulations are S8 and S9, for which
$\psi\approx 1.3$. Simulation S8 has $\alpha=0.26$, for which the
corresponding low amplitude simulations follow the linear predictions
for $\alpha_2$ (Eq.  (\ref{eq:prediction})).  Simulation S9 instead
has $\alpha=0.1$, for which the low amplitude simulation is already in
the saturated regime for $\alpha_2$. In both cases the diffusion
coefficient is found to be smaller than the value predicted from Eq.
(\ref{eq:prediction}) and we measure $f=0.75$ for S8 and $f=0.6$ for
S9. The results for simulation S8 thus show that increasing the warp
amplitude leads to a marginal reduction of the diffusion coefficient.
On the other hand. In the case of simulation S9, with $\alpha = 0.1$
for which a low amplitude warp is already enough to induce a reduction
of $\alpha_2$ with respect to linear theory, a further increase of
$\psi$ does not lead to a significant further reduction of $\alpha_2$.

These large amplitude simulations are also interesting because they reveal the
effects of the warp diffusion term on the evolution of the disc surface
density.  Fig.  \ref{fig:S8} shows the evolution of simulation S8 (solid line)
along with the evolution of the simple diffusive model of Eq.
(\ref{eq:pringle}) (dashed line). Note how the model reproduces well the
inflection of the surface density close to the warp location. The inflection
is due to the effect of $\nu_2$ on the evolution of $\Sigma$, as mentioned
earlier, and shown in \citet{pringle92}.  In contrast with the previous low
amplitude simulations, in this case, where the warp amplitude is large, this
effect does play a role.  Such effect has been described clearly in
\citet{pringle92} and more recently in \citet{LP06}. Note that such
dissipative effects cannot be reproduced by a purely wave-like warp
propagation.

\begin{figure}
\centerline{\epsfig{figure=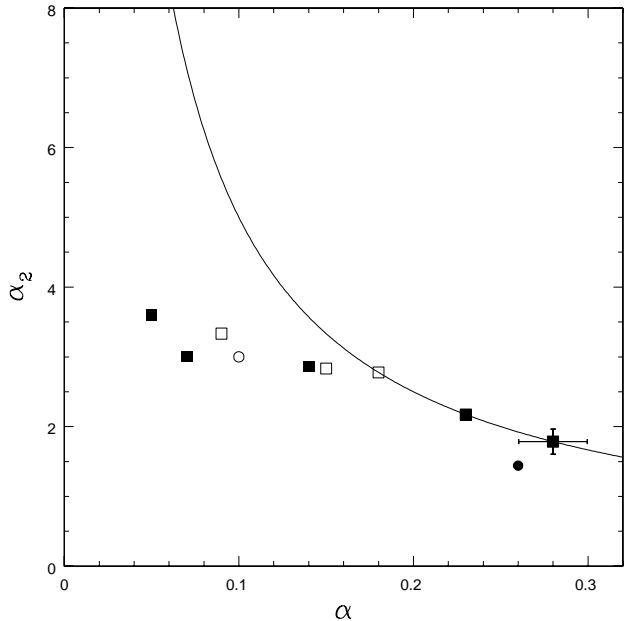,width=0.5\textwidth}}
\caption{Results of the numerical simulations. The points indicate the values
  of the diffusion coefficient $\alpha_2$ as a function of the viscosity
  coefficient $\alpha$. The solid symbols refer to simulations that do not use
  the `viscous switch' and should thus have a viscosity more closely
  approximating Navier-Stokes. The open symbols do use the viscous switch. The
  squares refer to the small warp amplitude case $\psi_{\rm max}\approx
  0.026$, while the circles refer to the large amplitude case $\psi_{\rm
    max}\approx 1.3$. The error bars shown represent the typical uncertainties
  on the diffusion coefficients. The solid line shows the expected value of
  $\alpha_2$ from the linear theory. It is evident that our simulations
  reproduce the expected results from linear theory for values of $\alpha >
  0.16$. Below this value we find that $\alpha_2$ appears to saturate at a
  value around $\alpha_{\rm max} \sim$ 3 -- 4.}
\label{fig:alpha2}
\end{figure}

\begin{figure*}
\centerline{\epsfig{figure=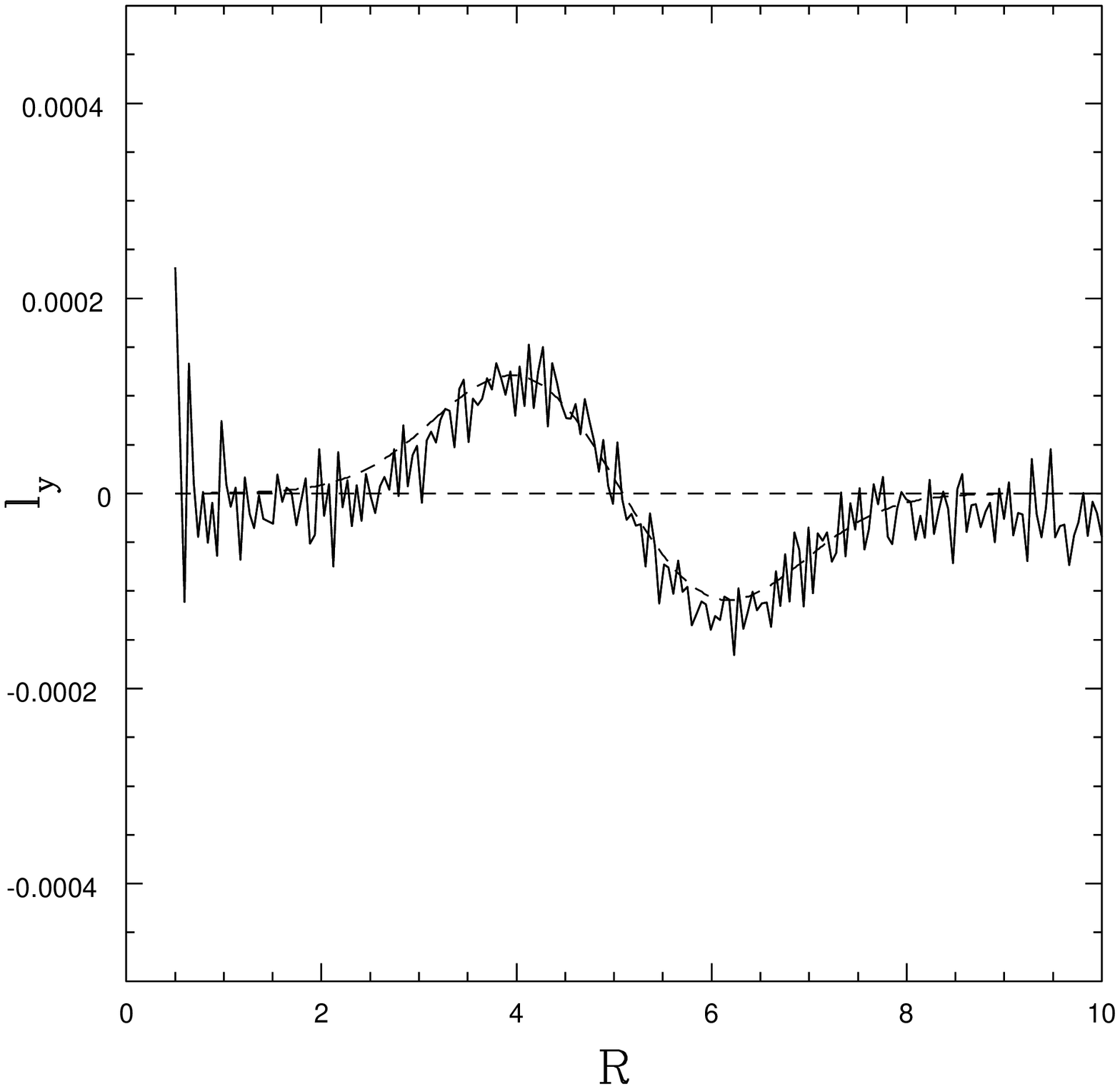,width=0.5\textwidth}
            \epsfig{figure=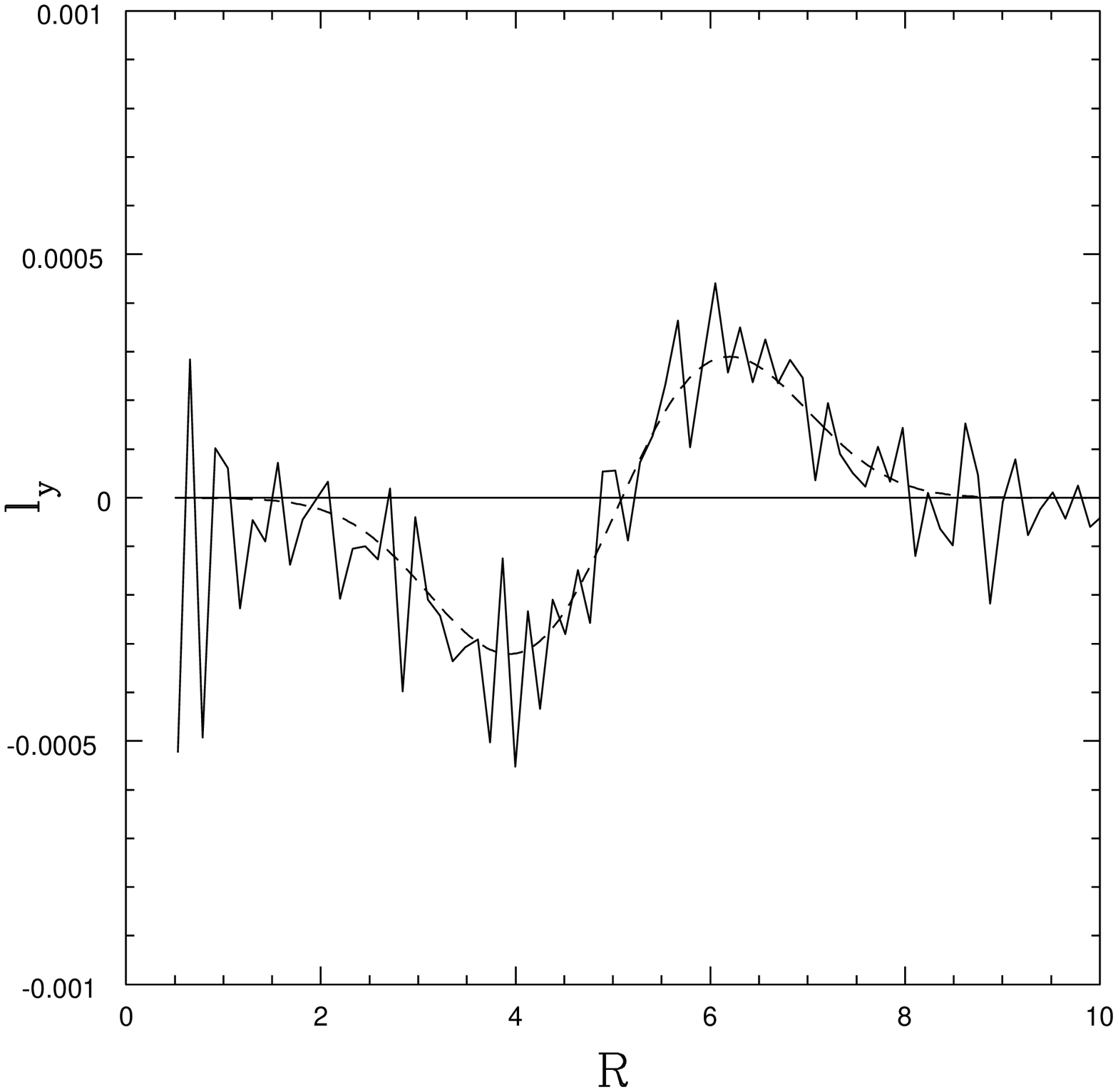,width=0.5\textwidth}}
          \caption{Evolution of the component $l_{y}$ of the unit vector
            ${\bf l}$. Initially $l_{y} = 0$ and it is the precessional
            torques, neglected in Equation (2), which cause $l_{y}$ to
            evolve.  Left panel: simulation S1b (with the viscous switch
            turned off).  Here the precession occurs in the direction
            predicted by the linear theory, but with a smaller precession
            rate.  The solid line represents the results of the SPH
            simulation, while the dashed line represents the result of a
            simple diffusion model including a precessional term, with
            $\alpha_3\approx 0.17$. The curves refer to $t=825$ (in units of
            the dynamical time at $R=1$).  Right panel: same plot for
            simulation S1, which has the same $\alpha$ and $\alpha_2$ as S1b,
            but the viscous switch is on. Here precession occurs in the
            opposite direction and the best diffusion/precession model able to
            reproduce the results has $\alpha_3\approx -0.46$. Here the curves
            refer to $t=840$ (in units of the dynamical time at $R=1$).  }
\label{fig:precession}
\end{figure*}

\subsection{The diffusion coefficients}

Each one of the high resolution simulations described in the previous
section is quite demanding in terms of computing time, and we could
only run a few such simulations. As a result, we could only obtain a
partial coverage of the relevant parameter space. Even so, we can
still draw some interesting conclusions from the limited available
data.

First we find that the evolution of the disc is well described by
Equation~(\ref{eq:pringle}), and the numerical results can be fit by
just varying the parameters $\alpha$ and $\alpha_2$. We stress that we
do not perform an actual statistical fit of the viscosity
coefficients, but simply choose them so as to match the evolution of
the numerical simulations. We estimate that the typical uncertainty on
these parameters is $\approx 7\%$ for $\alpha$ and $\approx 10\%$ for
$\alpha_2$.

Fig. \ref{fig:alpha2} shows the values of the diffusion coefficient for the
warp, as obtained from all of our thin disc simulations, for which propagation
of warp occurs diffusively. The solid symbols refer to the case where the
viscous switch is turned off (that should be closer to a Navier-Stokes
viscosity), while the open symbols refer to the case where the viscous switch
is on. For the two simulations with the highest $\alpha$, the point refers to
both cases, as the results were identical. The circles refer to the
high-amplitude simulations. As an example, we also show for one point the
typical uncertainties of the estimated parameters.

It can be seen that the points follow the expected relation $\alpha_2
= 1/2 \alpha$ at large $\alpha$, but as $\alpha$ decreases, the value
of $\alpha_2$ begins to deviate from the theoretical relation and
appears to saturate at a value of around $\alpha_{\rm max} \sim $3 --
4. However, perhaps surprisingly, comparison of runs S5 and S9 shows
that we do not observe any dependency of the saturation value on the
warp amplitude.

\subsection{Precession}
\label{sec:prec}

Our simulations also display some small precessional effects. This is
shown in Fig. \ref{fig:precession}, that shows the evolution of the
$y$-component of the angular momentum of the disc, $l_{y}(R)$.  This
is initially set to be zero (see Eq. (\ref{eq:initial})) but it then
grows due to internal precessional torques. Such torques are not
accounted for in the diffusion model by \citet{pringle92}, but they
are present in the full linear theory (e.g.  \citet{ogilvie99}).  In
interpreting our results we have added such terms in our simple
diffusion model, by adding a term on the right-hand side of equation
(\ref{eq:pringle}), in the form of (see \citealt{ogilvie99})
\begin{equation}
\left.\frac{\partial{\bf L}}{\partial t}\right |_{\rm prec} =  
\frac{1}{R}\frac{\partial}{\partial
R}\left(\nu_3R|{\bf L}| {\bf l}\times\frac{\partial{\bf l}}{\partial R}
\right),
\end{equation}
where we have introduced a third coefficient $\nu_3$ related to
precessional effects (with a corresponding $\alpha_3=\nu_3/\Omega
H^2$). The solid lines in Fig. \ref{fig:precession} show the evolution
of the SPH simulations, while the dashed lines refer to the simple
model (including precession). The two simulations shown in
Fig. \ref{fig:precession} are S1b (left panel) and S1 (right panel)
which have the same $\alpha$ and $\alpha_2$ but use a different
implementation of viscosity, so that S1 uses the viscous switch, while
S1b does not. The warp evolution of these two simulations agrees with
the predictions of linear theory for the evolution of the warp. They
do not, however, agree with linear theory with regard to disc
precession. Moreover, there is an obvious difference between the two
cases in that precession occurs in different directions. In order to
model the evolution, we therefore require a negative $\alpha_3$ term
in the case of S1 and a positive $\alpha_3$ for S1b. Note that the
linear analytic theory, using an {\it isotropic} Navier-Stokes
viscosity predicts a positive $\nu_3$ with magnitude given by
$\alpha_3=3/4$ \citep{pappringle83,ogilvie99}. The resulting values of
$\alpha_3$ for the simulations where we measure precession are shown
in Table \ref{tab:table}.  None of our simulations agrees with the
analytic results.

Note that in all cases $\alpha_3$ is much smaller than $\alpha_2$,
indicating that precession occurs on a much longer timescale than the
timescale for diffusion and/or damping of the warp. This implies that,
since internal precession is induced by the warp, it is unlikely to
play any role in the disc dynamics, because by the time it starts to
be important, the warp has already been diffused out, at least in the
case where there is no strong external torque to enforce a given
warped shape.

\section{Discussion}

\label{theory}

In this Section we discuss the outcome of our numerical simulations
and compare them with the predictions of the approximate analytic
theories.  Our general finding is that for values of $\alpha \ge
0.16$, the value of $\alpha_2$ follows the simple theoretical
prediction that $\alpha_2 = 1/2 \alpha$. However, for lower values of
$\alpha$ our simulations indicate that the value of $\alpha_2$ seems
to saturate at a upper limit of $\alpha_2 = \alpha_{\rm max} \sim $3
-- 4.

\subsection{The theory of the simple $\alpha_2 - \alpha$ relation}

In order to aid theoretical understanding of the processes involved, we first
introduce a simple, physical model to describe the diffusive evolution of a
warp in thin, viscous discs. This model essentially reproduces the more
rigorous analysis of \citet{pappringle83}, describing the main physical
processes in a simple and intuitive way.

\begin{figure}
\centerline{\epsfig{figure=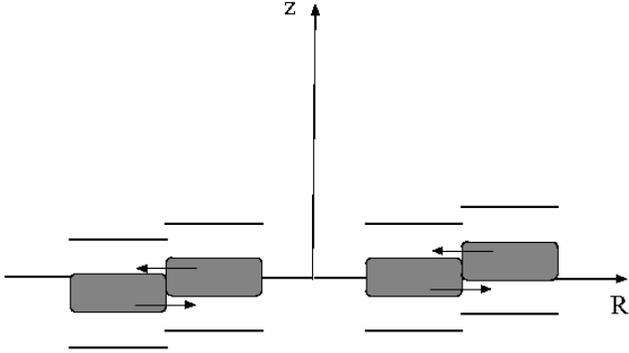,width=0.5\textwidth}}
\caption{Schematic view of a warped disc. The shaded areas indicate regions of
  higher pressure. The arrows indicate the direction of the horizontal
  pressure gradients induced by the warp. As a fluid element orbits around the
  centre, it feels an oscillating radial pressure gradient, whose amplitude is
  a linear function of the height $z$.}
\label{fig:scheme}
\end{figure}

The evolution of the warp is controlled essentially by hydrodynamical
processes taking place inside the disc. The main point is that a warp, coupled
with the vertical stratification (with typical scale-height $H$) in a thin
disc, induces a horizontal pressure gradient which depends on vertical height
$z$ (see Fig. \ref{fig:scheme}). Thus, as a fluid element orbits in the disc
with a frequency $\Omega$, it experiences a horizontal (radial) pressure
force, which oscillates at the same frequency $\Omega$. The crucial point is
that for a Keplerian disc this forcing term is resonant with the epicyclic
motion, since the epicyclic frequency $\kappa=\Omega$ for a Keplerian disc.
Hence the presence of the warp excites strong horizontal epicyclic motions
within the disc. The magnitude of the induced pressure gradient is given
roughly by
\begin{equation}
\frac{\partial p}{\partial R}\sim\frac{\partial p}{\partial z}\psi \sim
\frac{p\psi}{H}.
\end{equation}
Thus the corresponding force term in the horizontal momentum
equation is
\begin{equation}
\frac{1}{\rho}\frac{\partial p}{\partial R}\sim \frac{c_{\rm s}^2\psi}{H} \sim
H\Omega^2\psi,
\label{horizforce}
\end{equation}
where we have also used $p\sim c_{\rm s}^2\rho$. 

As is shown in more detail in Papaloizou \& Lin (1995) the forcing
term varies linearly with $z$, and therefore excites a response for
example in the radial velocity field which takes the approximate form
\begin{equation}
v_R = v_{\perp} \left( \frac{z}{H} \right) \cos \phi.
\label{VR}
\end{equation}
Here $\phi$ is the azimuthal angle in the disc, and $v_{\perp}$ denotes the
induced horizontal velocity in the radial direction at the disc `surfaces' $z
= \pm H$. If we let $x(t)$ be the radial Lagrangian displacement of a disc
particle at the disc surface relative to an unperturbed disc particle on a
circular orbit, then this implies that
\begin{equation}
x(t) = \left( \frac{v_{\perp}}{\Omega} \right) \sin \Omega t.
\label{xdot}
\end{equation}

Then the disc particle at position $x(t)$ is subject to three forces:
a restoring force which gives rise to the  epicyclic frequency $\kappa
= \Omega$, a damping force due to some form of viscous dissipation,
and a pressure force caused by the warp with amplitude given by
Equation~(\ref{horizforce}). Thus $x(t)$ obeys an equation of the form
\begin{equation}
\ddot{x}+\frac{\dot{x}}{\tau}+\Omega^2x = R\Omega^2\left(\frac{\psi H}{R}
\right)\cos{\Omega t}.
\label{eq:model}
\end{equation}
Here the amplitude of the forcing term on the r.h.s. is taken from
Equation~(\ref{horizforce}). The third term on the l.h.s. is the epicyclic
restoring force. The second term on the left-hand is a damping term, where we
have introduced a damping time-scale $\tau$, that describes the viscous
damping of the horizontal shear motion (described in Equation \ref{VR})
induced by the warp. If the horizontal shear motion is damped by a kinematic
viscosity $\nu_{z}$ then we have simply that
\begin{equation}
\tau\sim\frac{H^2}{\nu_{z}}
\label{eq:nuz}
\end{equation}
(Note that the typical lengthscale of the horizontal shear is $H$, rather than
$R$, hence the difference between the above formula and the corresponding one
associated with the viscous timescale in a differentially rotating disc,
$t_{\rm visc}\sim R^2/\nu_1$).

If we make the further assumption (which, as we have stressed, is not
necessarily true for a disc) that the viscosity is isotropic, so that
$\nu_{z}=\nu_1$, we obtain
\begin{equation}
\tau\sim\frac{H^2}{\nu_1}=\frac{1}{\alpha\Omega}.
\label{isotau}
\end{equation}

In line with our expectations (Equation~\ref{xdot}) we look for a
solution to Equation (\ref{eq:model}) in the form
\begin{equation}
x(t)=a\sin{\Omega t},
\label{eq:x}
\end{equation}
where now $a$ is the amplitude of the horizontal motions at the disc
surfaces $z = \pm H$. Substituting (\ref{eq:x}) into (\ref{eq:model}),
we find a solution for the amplitude of the horizontal motion
$a$
\begin{equation}
a=\psi H(\Omega\tau),
\end{equation}
which corresponds to a horizontal velocity of
\begin{equation}
v_{\perp} = a \Omega = \psi c_{\rm s} (\Omega \tau).
\end{equation}
The amplitude of the horizontal shear motion, $dv_R/dz$, is then given by
\begin{equation}
S=\frac{\Omega a}{H}= \Omega^2\tau\psi.
\end{equation}

If we add the assumption that the horizontal shear is damped by an {\it
  isotropic} viscosity with magnitude measured by the usual $\alpha$, this,
using Equation~(\ref{isotau}), then implies that
\begin{equation}
a \sim \frac{\psi}{\alpha}H,
\end{equation}
and the horizontal velocity is then given by~\footnote{As an aside, we
  emphasise that this estimate is not valid for small values of $\alpha$,
  because for $\alpha < \alpha_{\rm c}$ the propagation becomes wavelike, and
  in that case $v_\perp \sim (\psi/\alpha_{\rm c})c_{\rm s}$.}
\begin{equation}
v_{\perp}=\frac{\psi}{\alpha} c_{\rm s}.
\label{vperpiso}
\end{equation}

With this assumption the amplitude of the vertical shear, $S =
dv_R/dz$, is given by
\begin{equation}
\label{eq:shear}
S=\frac{\Omega a}{H}= \frac{\psi}{\alpha}\Omega.
\end{equation}

In order to relate the above description with the model described by Equation
(\ref{eq:pringle}), we note that in that equation the coefficient $\nu_2$ acts
formally on the radial derivative of the vertical component of the velocity
(since it dissipates the warp). By the argument described above, the
dissipation actually comes about because of damping of the resonantly induced,
horizontal, shear motion, that communicates the warp between different radii.
The definition of $\nu_2$ inherently implies that the rate at which energy is
dissipated in the two processes must therefore be the same. Thus we must have
\begin{equation}
\nu_2\left \langle \frac{\mbox{d}v_{z}}{\mbox{d}R}\right
\rangle^2=\nu_{z}\left \langle \frac{\mbox{d}v_R}{\mbox{d}z}\right \rangle^2,
\label{eq:dissipation}
\end{equation} 
where angled brackets imply suitable vertical and azimuthal averages.

Now, in a warped disc $v_{z}=\psi R\Omega$ and $\mbox{d}v_{z}/\mbox{d}
R=\psi\Omega$, which, in equation (\ref{eq:dissipation}) gives
\begin{equation}
\nu_2= \nu_{z}(\Omega\tau)^2=  \Omega H^2(\Omega\tau),
\end{equation}
where, in the last equality, we have used the definition of $\nu_{z}$,
Equation (\ref{eq:nuz}). This is equivalent to writing
\begin{equation}
\alpha_2 =  \Omega \tau.
\label{alpha2}
\end{equation}

If the damping time-scale $\tau$ is indeed given by isotropic viscous
dissipation, then, as above, we have $\tau\Omega=1/\alpha$ and we
finally recover the desired scaling of $\nu_2$ with $\alpha$ in the
linear case (cf.  equation (\ref{eq:alpha2}))
\begin{equation}
\frac{\nu_2}{\nu_1} \sim \frac{1}{\alpha^2}.
\end{equation}

It is worth noting here that this result is somewhat counter-intuitive, in
that it implies that the damping of the warp occurs {\it more rapidly} when
the viscosity is smaller! This comes about because a small value of isotropic
viscosity (small $\alpha$) permits a large resonant velocity ($v_R \propto
1/\alpha$). The dissipation rate depends on $\nu v_R^2 \propto 1/\alpha$, and
so {\it increases} as $\alpha$ decreases. It is evident that the analysis must
break down at some stage as $\alpha \rightarrow 0$.

\subsection{Comparision with the simulations}

Most of the results shown in Figure \ref{fig:alpha2} come from a set
of simulations all with $H/R = 0.01333$ and $\psi_{\rm max}=0.026$ but
with varying values of $\alpha$. For these simulations we find that
relationship between $\alpha$ and $\alpha_2$ predicted by simple
linear theory holds for values of $\alpha$ above around $\alpha \approx
0.16$ where $\alpha_2 \approx 3$.  Below this critical value of
$\alpha$, $\alpha_2$ remains approximately constant.

We now need to ask what mechanism(s) might give rise to such an
effect, and in addition whether they correspond to some valid physics,
or are merely numerical artifacts.

We have noted that from simple physical considerations we expect that
$\alpha_2 \approx \Omega \tau$, where $\tau$ is the timescale for damping the
warp-induced horizontal shearing motions. Thus an upper limit to $\alpha_2$
can come about through there being an upper limit to the damping timescale
$\tau$; that is, we need to identify a mechanism which prevents the damping
taking place too slowly. For isotropic viscosity, the damping timescale is
such that $\Omega \tau = 1/\alpha$, so that the lower the viscosity, the
longer the timescale. Thus, paradoxically, in order to set an upper limit on
$\alpha_2$, i.e. in order to stop the warp being dissipated too fast, we need
to find a mechanism which gives rise to {\it increased} dissipation. If there
is such a process such that $\tau < K \Omega^{-1}$, where $K$ is some
constant, then we would find that $\alpha_2 < K = \alpha_{\rm max}$, in line
with our findings.

So what physical or numerical process(es) might give rise to enhanced
dissipation for small values of $\alpha$?

We have noted that for isotropic viscosity the Mach number of the horizontal
velocity induced by the warp is (to within factors of order unity)
\begin{equation}
v_\perp/c_{\rm s} = {\cal{M}} = S/\Omega = \psi/\alpha.
\label{eq:psialpha}
\end{equation}
The above equation shows that the horizontal resonant motion becomes
supersonic for $\alpha \le \psi$.  In our simulations breakdown of the simple
relation occurs for $\alpha \approx 0.16$ and $\psi \approx 0.026$, and
therefore when ${\cal{M}} \approx 0.16$. Thus, breakdown occurs when the
velocity difference across the disc thickness $|v_{\perp}(H) - v_{\perp}(-H)|
\approx 0.32 c_s$

\begin{figure*}
\centerline{\epsfig{figure=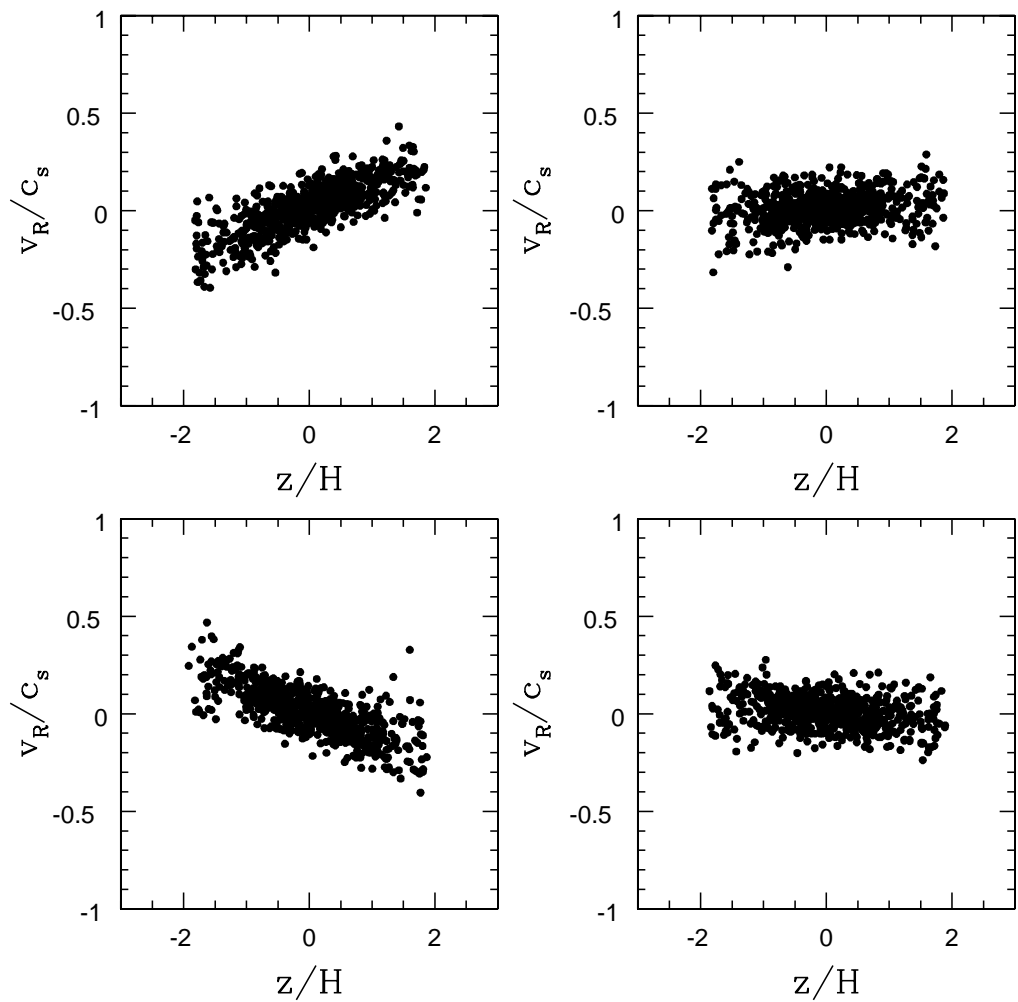,width=0.8\textwidth}}
\caption{Run S6. The radial Mach number induced by the warp as a function of
  height above the disc at $R=5$ and $\phi=0$ (upper left), $\phi=\pi/2$
  (upper right), $\phi=\pi$ (lower left) and $\phi=3\pi/2$ (lower right). Here
  the cylindrical coordinate system is such that the $z$-axis lies along the
  local direction of the angular momentum vector ${\bf l}$.  The disc
  scaleheight here is $H = 0.0665$. The amplitude of this flow at $z = \pm H$
  is ${\cal M} \approx 0.1$. }
\label{fig:osc}
\end{figure*}

The radial shear flow induced by the warp can be clearly noted in Fig.
\ref{fig:osc}, which displays the radial Mach number for simulation S6 at
$t=195$, at $R=5$ and at different azimuthal positions, that is $\phi=0$
(upper left), $\phi=\pi/2$ (upper right), $\phi=\pi$ (lower left) and
$\phi=3\pi/2$ (lower right) (Here, we use a cylindrical coordinate system such
that the $z$-axis lies along the local direction of the angular momentum
vector ${\bf l}$). At height $z = \pm H$ the amplitude of this radial
oscillation is $v_{\perp}/c_{\rm s}\approx 0.1$, much less than the value of
0.37 expected for this simulation based on Eq. (\ref{eq:psialpha}).

We have also run two simulations (S8 and S9) with a much larger warp
amplitude, that is $\psi_{\rm max} = 1.3$. For simulation S9, we see that
$\alpha\approx 0.1$, in which case we expect the Mach number of the radial
resonant motion to be of the order of ${\cal{M}} \approx 13$, which is much
larger than the Mach number ${\cal{M}} \approx 0.16$, above which we have
observed the standard relation between $\alpha$ and $\alpha_2$ to break down.
This simulation should be compared with simulation S5, which has a very
similar $\alpha$ and a very similar $\alpha_2$, but a much smaller $\psi$ and
therefore a smaller predicted Mach number of ${\cal{M}}\approx 0.29$ (but
still large enough to be in the saturated regime). We then see that increasing
further the warp amplitude does not produce a significant decrease in
$\alpha_2$. Thus although for a larger warp the amount of dissipation
increases, the dissipation timescale appears to remain unchanged at
$\tau\approx 3\Omega^{-1}$. In the case of simulation S8, instead, we have
$\alpha\approx 0.26$, so that the predicted Mach number would be
${\cal{M}}\approx 5$. The corresponding low amplitude simulation (S0), which
has $\alpha\approx 0.28$ and predicted ${\cal{M}}\approx 0.1$, lies in the
non-saturated regime, and in this case increasing the warp amplitude does
result in an increased dissipation and a reduction in the value of $\alpha_2$,
which for S8 lies marginally below the predicted relation.

\begin{figure*}
\centerline{\epsfig{figure=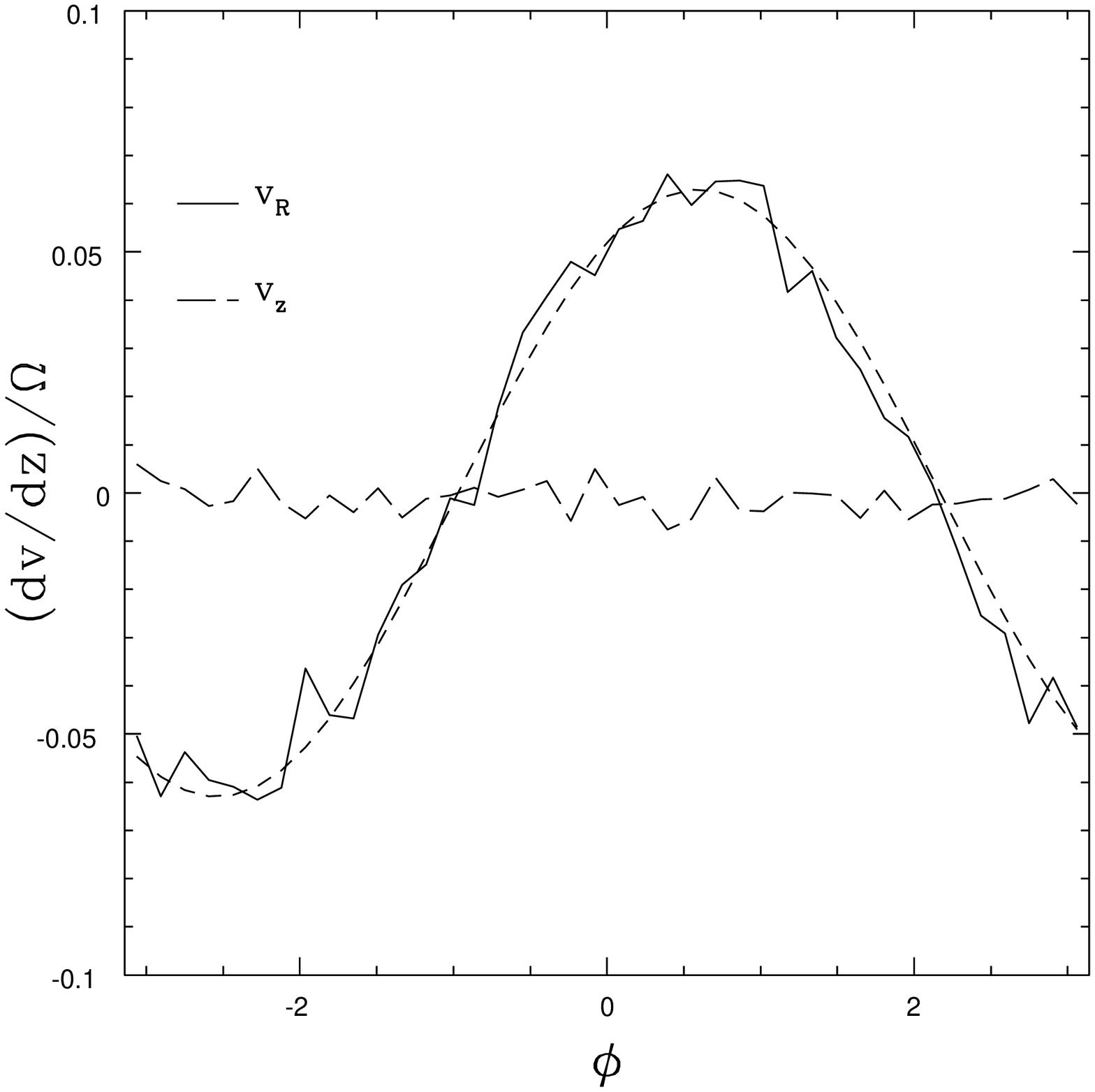,width=0.5\textwidth}
            \epsfig{figure=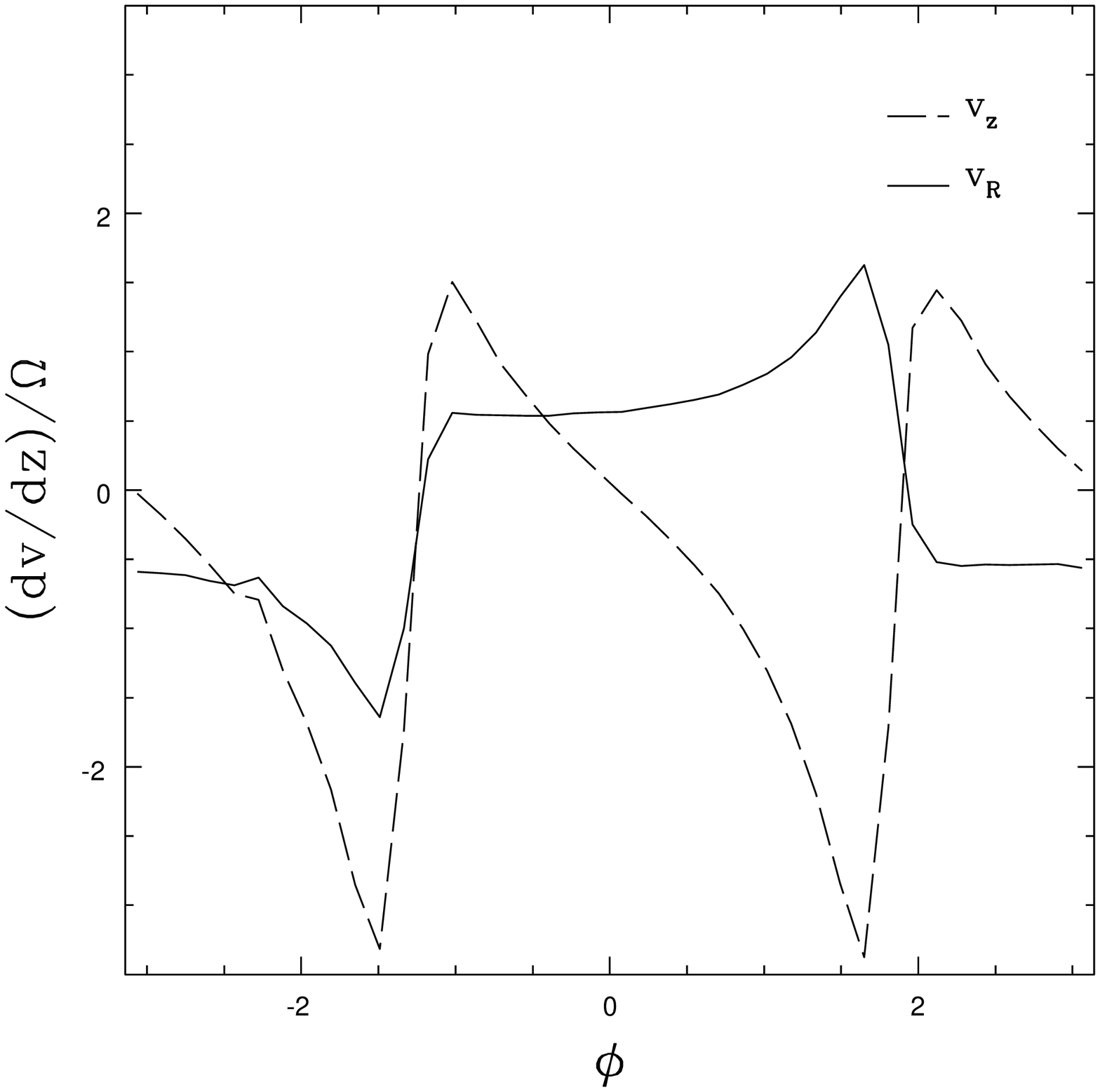,width=0.5\textwidth}}
          \caption{Vertical gradient of the resonant flow induced by the warp
            for simulations S0b, for which $\psi\approx 0.026$ (left) and S9
            (right), for which $\psi\approx 1.3$. The linear theory predicts
            to first order in $\psi$ that $\de v_R/\de z \propto \psi \sin
            (\phi - \phi_0)$ and $\de v_z/\de z \propto \psi^2 \sin 2(\phi -
            \phi_0)$. The solid line shows $\mbox{d}v_R/\mbox{d}z$, while the
            long-dashed line shows $\mbox{d}v_z/\mbox{d}z$. The short-dashed
            line is a sinusoidal fit to $\mbox{d}v_R/\mbox{d}z$ for S0b. Here
            the cylindrical coordinate system is such that the $z$-axis lies
            along the local direction of the angular momentum vector ${\bf
              l}$.}
\label{fig:shear}
\end{figure*}

\subsection{High Mach number and enhanced dissipation}

We have established that the results of our simulations can be understood if
the internal disc flows, induced by the warp, result in higher dissipation
than would be predicted by linear theory (assuming Navier-Stokes viscosity)
when the Mach number of the flows exceeds a value of around ${\cal{M}} \sim
0.16$, i.e. when the relative flow velocity across a full disc thickness
exceeds around $0.32 c_s$. The internal flows predicted by the analytic theory
are in fact of two types (see, for example, Equations (102) and (103) of
Ogilvie, 1999). The first is the sinusoidally oscillating, linear shear flow
discussed in Section~\ref{theory} above\footnote{This radial ($v_R$) flow is
  part of a resonant epicyclic motion (see Section 4.1). It is therefore
  accompanied by an azimuthal flow of the form $v_{\phi} \propto z$, with
  amplitude $v_R = 2 v_{\phi}$.}, which takes the form $v_R \propto z$ and
oscillates with frequency $\Omega$. This flow has an amplitude which scales
linearly with $\psi$, and in the linear theory is indeed proportional to
$\psi/\alpha$. The second, whose magnitude depends more strongly on the size
of the warp (that is, it scales with $\psi^2$), is a sinusoidally pulsating,
homologous flow, which is alternately expanding and contracting with angular
frequency $2 \Omega$ and is of the form $v_z \propto z$. We have calculated
the vertical gradients of these two types of flows by fitting a straight line
to the velocity data, such as those shown in Fig. \ref{fig:osc}. We plot in
Fig. \ref{fig:shear} the results of this procedure as a function of azimuth
for simulations S0b (left) and S9 (right). The typical uncertainty for these
data points ranges from 5 to 10 per cent. In both plots the solid line refers
to the gradient of $v_{R}$, while the long-dashed line refers to the gradient
of $v_z$. In the S0b case, the short-dashed line is a sinusoidal fit to the
data points for $v_R$, with an amplitude equal to 0.063.

Let us first consider the case of simulation S0b. As discussed above, this
simulation agrees well with the linear theory and with its prediction for the
warp diffusion coefficient. In agreement with linear theory, for this low
amplitude warp, we find that the induced resonant flow is dominated by the
radial component, the vertical one being negligible. The radial flow is indeed
oscillating with frequency $\Omega$ as predicted by Ogilvie (1999). We have
shown above (Eq.  (\ref{eq:shear})) that the amplitude of this oscillation
should be (within factors of order unity) of the order of $\psi/\alpha
\sim 0.093$ for S0b.  Based on the results shown in Fig. \ref{fig:shear} we
can estimate the exact proportionality factor to be
\begin{equation}
\frac{\de v_R}{\de z}\approx 0.65\frac{\psi}{\alpha}\Omega.
\end{equation}
We can now check the predicted scaling with $\psi/\alpha$ as we decrease the
value of $\alpha$. To this end, we have repeated the above analysis for S4 and
S6, and we have found in both cases a roughly sinusoidal oscillation of
$(\mbox{d}v_R/\mbox{d}z)/\Omega$, with amplitude equal to $0.1=0.54
\psi/\alpha$ for S4 and to $ 0.12=0.33 \psi/\alpha$ for S6. Hence, we do not
confirm the predicted scaling, the value of $\mbox{d}v_R/\mbox{d}z$ for S6
being a half of its predicted value. This is further evidence for enhanced
dissipation.  Note that this enhanced dissipation is only found for small
values of $\alpha$ and it appears to increase as we increase the value of
$\psi/\alpha$.

The presence of both the radial and the vertical flow can be seen clearly in
the right panel of Fig. \ref{fig:shear} and in Figure 13 which corresponds to
the run S9 which has a strong warp $\psi \sim 1$. The structure of this
pulsating flow is shown in Figure 14, where we show the Mach number in the
$z$-direction as a function of $z$ at various azimuths in the disc (as before,
the cylindrical coordinate system used here is such that the $z$-axis lies
along the local direction of ${\bf l}$). As can be seen, the flow is highly
supersonic, with disc elements approaching each other at velocities up to
$\approx 10 c_s$. As shown by Ogilvie (1999) these flows are formal solutions
to the analytic equations, even for large values of $\psi$, when the Mach
numbers of the flows are expected to be large. Ogilvie (1999) does not,
however, discuss the stability of such flows. Note that while both these flow
appear to have the correct azimuthal periodicity (with frequency $\Omega$ for
the radial component and $2\Omega$ for the vertical one), they are not
sinusoidal as additional Fourier components become important for large $\psi$
(Ogilvie 1999, Appendix). Note, in particular, the large amplitude and the
sharp jumps of the vertical compression. Across a very narrow azimuthal range
(and therefore over a timescale much shorter then dynamical) the velocity
changes by a factor much larger than the sound speed.  Given the short
timescale, this change can be only due to pressure forces and artificial
viscosity in this strongly compressional motion. Note also that the negative
amplitude of the compressional motion is larger than the positive one,
indicating that a large amount of kinetic energy is dissipated in the process.
This also implies that the pulsating flow is not of the form $v_z\propto z$ at
all times.

\begin{figure}
\centerline{\epsfig{figure=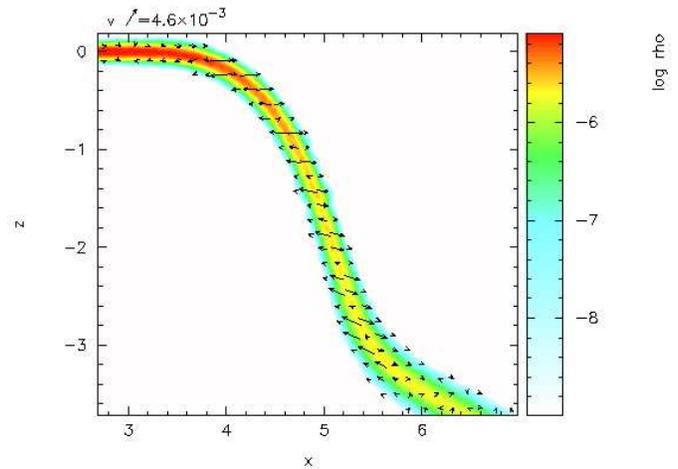,width=0.5\textwidth}}
  \caption{Colour plot of the vertical density structure of simultion S9, at
    $t=185$, close to the warp location at $R=5$. The arrows indicate the
    velocity structure. The vertically shearing radial flow induced by the
    warp has, in this large warp amplitude case, a significant component along
    the local vertical direction, determining a strong compression and
    rarefaction of the disc.}
\label{fig:flow2}
\end{figure}

\subsubsection{A numerical explanation?}

One possibility we need to address is whether the enhanced dissipation
that we find in our simulations at high Mach number is a purely
numerical effect. From the simulations presented here it is hard to be
certain that it is not. But it is not easy to see what the effect
might be.

It is certainly true that the flows are only moderately well resolved
in our simulations: in order to treat discs which are thin enough that
warp propagation occurs diffusively we were only able to consider
simulations in which the average number of SPH smoothing lengths
across the thickness of the disc is around 3.3. But the explanation
cannot simply be due to lack of resolution, since our simulations
agree with the analytic predictions for $\alpha_2$ for the larger
values of $\alpha$, in the regime where the induced velocities are
low, and analytic solutions are expected to be a good
approximation. Disagreement only occurs once the Mach number of the
induced shear flow increases to about 0.16 across a scaleheight $H$,
or to about 0.096 across a smoothing length $h$. In addition we obtain
similar results for $\alpha_2$ whether the SPH viscous switch is
turned on or off. Thus it seems to us unlikely that either lack of
resolution or the nature of the SPH stress tensor is the cause of the
disagreement.

It is evident, however, that the disagreement between the approximate
analytic theory and the numerical simulations in respect of precession
rates does depend on the nature of the numerical viscosity. Our
finding that none of our simulations confirms the analytic
predictions, and that even for the same formal $\alpha$ we obtain
opposite signs for the precession rates depending on whether or not
the viscous switch is on or off, implies that the induced precession
is very sensitive to the precise details of disc viscosity. This
implies that if the viscous stress tensor within the disc is not
precisely of the Navier-Stokes form, and as we have remarked above it
is very unlikely that it is, then the analytic predictions for the
precession rates are unlikely to be relevant to real accretion
discs. Luckily, as we have remarked above, such precession is in
general only a minor effect.

\begin{figure*}
\centerline{\epsfig{figure=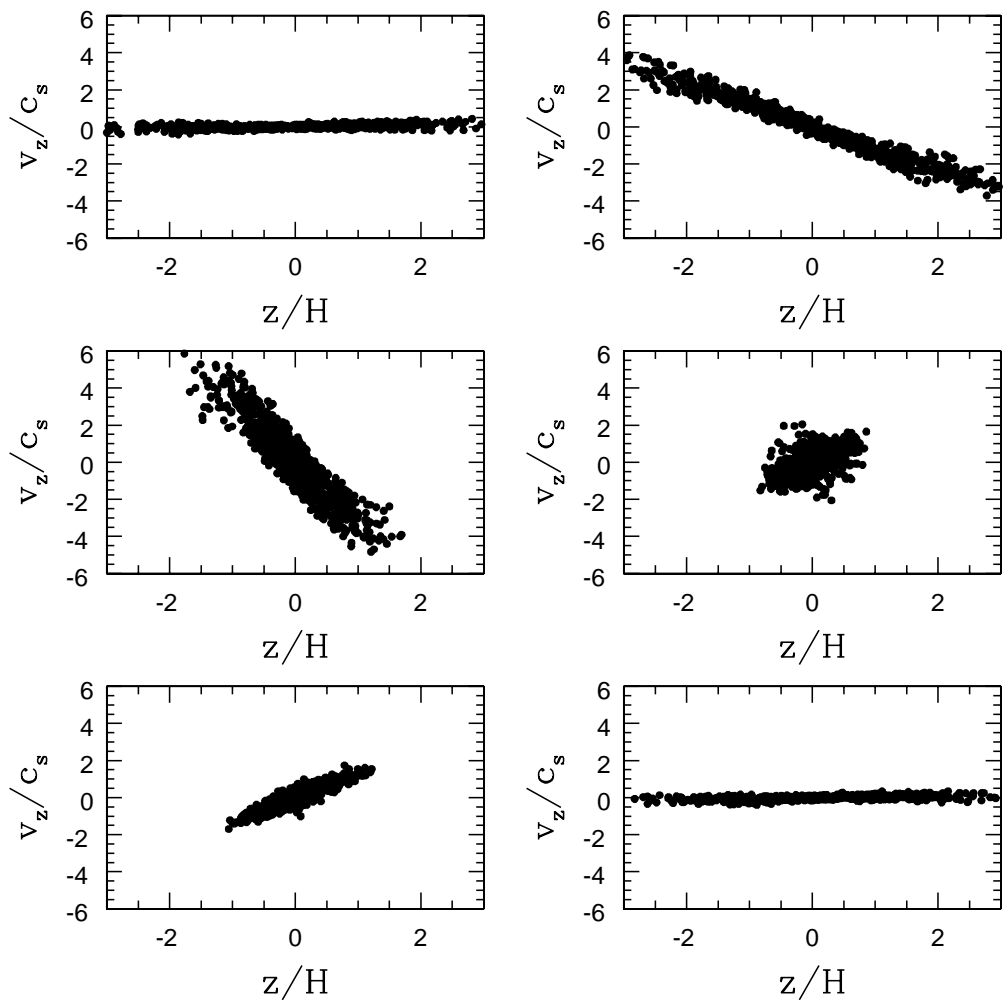,width=0.8\textwidth}}
\caption{Velocity along the local direction of the angular momentum vector, in
  units of the local sound speed, for simulation S9, at $t=185$, $R=5$ and
  $\phi=0$ (upper left), $\phi=0.3\pi$ (upper right), $\phi=\pi/2$ (middle
  left), $\phi=0.6\pi$ (middle right), $\phi=0.7\pi$ (lower left) and at
  $\phi=\pi$ (lower right). Here, the coordinate system is such that the
  $z$-axis lies along the local direction of the angular momentum ${\bf l}$.
  Between $\phi=0$ and $\phi=\pi/2$ the disc undergoes supersonic compression
  ($dv_z/dz <0$), with a Mach number at $z= \pm H = \pm 0.0665$ approaching 3.
  Between $\phi=0.5 \pi$ and $\phi=0.6 \pi$ these velocities are abruptly
  halted. At later phases the disc undergoes supersonic rarefaction
  ($dv_z/dz>0$), until it reaches its original structure at $\phi = \pi$. The
  sequence is repeated between $\phi = \pi$ and $\phi = 2 \pi$ because the
  orbiting disc gas has to receive a vertical velocity nudge twice per orbit
  in order to prevent the orbits of gaseous disc elements from crossing.}
\label{fig:comp}
\end{figure*}

\subsubsection{A physical explanation?}

When the Mach numbers of the internally induced flows approach unity
it seems physically plausible that such flows might become
unstable. If such instabilities gave rise to enhanced dissipation,
then they could provide an explanation for our results. 

Instabilities in a steady shear flow typically grow at a rate $\gamma \sim S$,
the rate of shear (e.g. \citealt{drazin81}). In an oscillating shear flow one
might expect instability to be suppressed unless the growth rate exceeds the
rate of oscillation i.e. unless $S > \Omega$.  However, Gammie et al (2000)
demonstrate that shear instabilities, caused by parametrically unstable modes,
can exist with $\gamma \sim S$ even when $S < \Omega$ (and in particular they
predict instability for $S/\Omega\sim {\cal{M}}> 30\alpha$). For the largest
scale modes of size $\sim H$ the damping timescale would be $\sim H^2/\nu \sim
\tau \sim (\alpha \Omega)^{-1}$. Thus we might expect shear instabilities to
be able to grow when $\gamma \tau > 1$, that is when $\psi > \alpha^2$ or when
${\cal{M}} > \alpha$. Using these arguments, for the set of simulations with
$\psi = 0.026$, we would then expect shear instabilities to occur when $\alpha
< 0.16$, which is in agreement with our findings. To discover whether this
agreement is fortuitous or not will require further work.

We have not demonstrated, however, that a shear flow becomes unstable
simply because it is supersonic, and to this extent the arguments we
put forward here are speculative. Indeed one could argue in this case
that the particles on the top and bottom of the disc are essentially
following free, elliptical trajectories around a central point
mass. However, this is not strictly true because fluid elements
separated vertically by $\sim H$ do need to interact, and to receive a
vertical velocity nudge of order $\sim c_{\rm s}$ twice per orbit,
i.e. on a timescale $\sim (1/2) \Omega^{-1}$, in order to remain a
distance $\sim H$ apart.  Once the horizontal fluid motions become
supersonic, i.e. once $S \sim \Omega$, the fluid elements separated
vertically by $\sim H$ can no longer maintain pressure
communication. Then unless the flow adheres exactly to the simple
solution found in analytic theory, without any perturbations, then the
required velocity nudge has to be communicated through shocks.

Indeed it is this vertical velocity nudge, required twice per orbit,
which gives rise to the compressive, pulsatory flow of the form $v_z
\propto z$ with frequency $2 \Omega$. Again for this motion, unless
the flow adheres exactly to the analaytical solution, and there are no
perturbations present within the disc, then once the pulsatory motions
approach sonic velocities, the pulsations must result in shock
formation and enhanced dissipation.

\section{Conclusions}

Using numerical simulations we have explored the propagation and damping of
warped accretion discs in the regime ($\alpha \geq \alpha_{\rm c}$) when
propagation occurs diffusively. In this regime we find that warp propagation
is described reasonably well by the simple equation given by Pringle (1992),
which involves just two viscosity parameters $\alpha$ and $\alpha_2$. This is
true, and gives the same $\alpha_2 - \alpha$ relationship, whether or not the
SPH `viscous switch' is turned on. As expected from linear analysis, and from
the simple physical arguments presented in Section~4, we find that as $\alpha$
decreases, $\alpha_2$ increases and at large values of $\alpha$ our
simulations agree with the predictions of previous analytic work. This is the
first time that this has been demonstrated numerically.  We also find,
however, that, if $\alpha$ decreases below some value $\alpha_{\rm sat}
\approx 0.16$, enhanced dissipation occurs which acts to damp the warp-driven,
internal, resonant motions, and so prevents a further rise in $\alpha_2$. We
find a maximum value of $\alpha_2 = \alpha_{\rm max} \approx$ 3 -- 4. We
speculate that this maximum value comes about because at low values of
$\alpha$ the induced horizontal shearing motions become close to sonic, and
are therefore subject to instabilities and enhanced dissipation. Further
numerical work is required to test these ideas.

In line with the expectations of linear theory we also find that the
warp induces precession which causes the disc to twist. As predicted,
the precession timescales are typically smaller than the timescales
for warp propagation and damping. However, the exact values of the
precession rates, and even the direction of precession, are not in
agreement with the predictions of linear theory. This presumably
reflects at some level the degree to which SPH viscosity fails to
provide an accurate model of isotropic Navier-Stokes viscosity, which
is assumed in the linear calculations. Indeed it is for the
simulations with the `viscous switch' turned on (so that dissipation
only occurs in regions of converging flow), for which the SPH
viscosity can be expected to be least like Navier-Stokes, that the
direction of precession is the opposite to that predicted by linear
theory. Since the viscosity in real accretion discs is unlikely to be
Navier-Stokes, this implies that theoretical predictions of
warp-induced precession rates need to be treated with caution.

The numerical method we have used is SPH. This has the advantage of
being a Lagrangian code, and so can deal straightforwardly with thin
warped discs, without the need of careful treatment of strong density
contrasts across neighbouring grid cells. It has, however, the
disadvantage that the fluid viscosity ($\alpha$) depends on particle
density, which therefore implies that varying viscosity involves
varying numerical resolution. As can be seen from Table~1, even using
one or two million particles, for the parameter ranges we are able to
investigate the number of smoothing lengths per disc scaleheight is
never very large, varying from around 10 for the thickest disc down to
1.7 for the thinnest disc. It is further worth noting that although we
do find evidence of enhanced dissipation as $\alpha$ decreases,
i.e. as the amplitude of the internal resonant shearing motions
increases, it is not clear that at this resolution the fluid dynamical
effects which give rise to such enhanced dissipation (e.g. fluid
instabilities, shocks, etc.) are being captured correctly. Thus the
values we find for $\alpha_{\rm max}$ need to be treated with
appropriate caution. In addition we find that the small precessional
effects induced by the warp depend sensitively on the exact
implementation of the SPH viscosity. It would be instructive to
compare the results we obtain here with results obtained using a grid
based numerical technique which would be better able to deal with
shock-capturing, and thus better able to identify the source of the
enhanced dissipation we find at large amplitudes of the horizontal
shear. It is also important to incorporate magnetic effects, since it
is these which are thought to be the basic cause of disc `viscosity',
and it seems likely that these may not be well modeled by the
assumption of an isotropic Navier-Stokes viscosity.

\section*{Acknowledgments}

We thank Gordon Ogilvie for useful advice which led to a redrafting of an
earlier version of this paper, the anonymous Referee for a constructive report
and Steve Lubow for interesting discussions. The simulations described in this
paper have been performed at the UK Astrophysical Fluid Facility (UKAFF). The
visualization of the simulations shown in Fig. \ref{fig:flow2} has been made
using the SPLASH software by Dan Price.

\bibliographystyle{mn2e} \bibliography{lodato}

\begin{thebibliography}{}

\bibitem[\protect\citeauthoryear{Artymowicz \& Lubow}{Artymowicz \&
  Lubow}{1994}]{lubow94}
Artymowicz P.,  Lubow S.~H.,  1994, ApJ, 421, 651

\bibitem[\protect\citeauthoryear{Balbus \& Hawley}{Balbus \&
  Hawley}{1998}]{balbusreview}
Balbus S.~A.,  Hawley J.~F.,  1998, Reviews of Modern Physics, 70, 1

\bibitem[\protect\citeauthoryear{{Bardeen} \& {Petterson}}{{Bardeen} \&
  {Petterson}}{1975}]{bardeen75}
{Bardeen} J.~M.,  {Petterson} J.~A.,  1975, ApJ, 195, L65

\bibitem[\protect\citeauthoryear{Bate, Bonnell \& Price}{Bate
  et~al.}{1995}]{bate95}
Bate M.~R.,  Bonnell I.~A.,    Price N.~M.,  1995, MNRAS, 277, 362

\bibitem[\protect\citeauthoryear{{Begelman}, {King} \& {Pringle}}{{Begelman}
  et~al.}{2006}]{begelman06b}
{Begelman} M.~C.,  {King} A.~R.,    {Pringle} J.~E.,  2006, MNRAS, 370, 399

\bibitem[\protect\citeauthoryear{Benz}{Benz}{1990}]{benz90}
Benz W.,  1990, in Buchler J.,  ed., The Numerical Modeling of Nonlinear
  Stellar Pulsations Kluwer, Dordrecht

\bibitem[\protect\citeauthoryear{{Chiang} \& {Murray-Clay}}{{Chiang} \&
  {Murray-Clay}}{2004}]{chiang04}
{Chiang} E.~I.,  {Murray-Clay} R.~A.,  2004, ApJ, 607, 913

\bibitem[\protect\citeauthoryear{{Coleman} \& {Dopita}}{{Coleman} \&
  {Dopita}}{1992}]{coleman92}
{Coleman} C.~S.,  {Dopita} M.~A.,  1992, Proceedings of the Astronomical
  Society of Australia, 10, 107

\bibitem[\protect\citeauthoryear{{Drazin} \& {Reid}}{{Drazin} \&
  {Reid}}{1981}]{drazin81}
{Drazin} P.~G.,  {Reid} W.~H.,  1981, NASA STI/Recon Technical Report A, 82,
  17950

\bibitem[\protect\citeauthoryear{{Fromang} \& {Papaloizou}}{{Fromang} \&
  {Papaloizou}}{2007}]{fromang07}
{Fromang} S.,  {Papaloizou} J.,  2007, A\&A, 468, 1

\bibitem[\protect\citeauthoryear{{Gammie}, {Goodman} \& {Ogilvie}}{{Gammie}
  et~al.}{2000}]{gammie00}
{Gammie} C.~F.,  {Goodman} J.,    {Ogilvie} G.~I.,  2000, MNRAS, 318, 1005

\bibitem[\protect\citeauthoryear{{Gingold} \& {Monaghan}}{{Gingold} \&
  {Monaghan}}{1977}]{gingold77}
{Gingold} R.~A.,  {Monaghan} J.~J.,  1977, MNRAS, 181, 375

\bibitem[\protect\citeauthoryear{{Herrnstein}, {Greenhill} \&
  {Moran}}{{Herrnstein} et~al.}{1996}]{herrnstein96}
{Herrnstein} J.~R.,  {Greenhill} L.~J.,    {Moran} J.~M.,  1996, ApJ, 468, L17

\bibitem[\protect\citeauthoryear{{King} \& {Pringle}}{{King} \&
  {Pringle}}{2006}]{king06}
{King} A.~R.,  {Pringle} J.~E.,  2006, MNRAS, 373, L90

\bibitem[\protect\citeauthoryear{{King}, {Pringle} \& {Livio}}{{King}
  et~al.}{2007}]{king07}
{King} A.~R.,  {Pringle} J.~E.,    {Livio} M.,  2007, MNRAS, 376, 1740

\bibitem[\protect\citeauthoryear{{Larwood}, {Nelson}, {Papaloizou} \&
  {Terquem}}{{Larwood} et~al.}{1996}]{larwood96}
{Larwood} J.~D.,  {Nelson} R.~P.,  {Papaloizou} J.~C.~B.,    {Terquem} C.,
  1996, MNRAS, 282, 597

\bibitem[\protect\citeauthoryear{{Lodato} \& {Pringle}}{{Lodato} \&
  {Pringle}}{2006}]{LP06}
{Lodato} G.,  {Pringle} J.~E.,  2006, MNRAS, 368, 1196

\bibitem[\protect\citeauthoryear{{Lubow} \& {Ogilvie}}{{Lubow} \&
  {Ogilvie}}{2000}]{lubow00}
{Lubow} S.~H.,  {Ogilvie} G.~I.,  2000, ApJ, 538, 326

\bibitem[\protect\citeauthoryear{{Lubow}, {Ogilvie} \& {Pringle}}{{Lubow}
  et~al.}{2002}]{lubow02}
{Lubow} S.~H.,  {Ogilvie} G.~I.,    {Pringle} J.~E.,  2002, MNRAS, 337, 706

\bibitem[\protect\citeauthoryear{{Lucy}}{{Lucy}}{1977}]{lucy77}
{Lucy} L.~B.,  1977, AJ, 82, 1013

\bibitem[\protect\citeauthoryear{Monaghan}{Monaghan}{1992}]{monaghan92}
Monaghan J.~J.,  1992, ARA\&A, 30, 543

\bibitem[\protect\citeauthoryear{Murray}{Murray}{1996}]{murray96}
Murray J.~R.,  1996, MNRAS, 279, 402

\bibitem[\protect\citeauthoryear{{Nelson} \& {Papaloizou}}{{Nelson} \&
  {Papaloizou}}{1999}]{nelson99}
{Nelson} R.~P.,  {Papaloizou} J.~C.~B.,  1999, MNRAS, 309, 929

\bibitem[\protect\citeauthoryear{{Nelson} \& {Papaloizou}}{{Nelson} \&
  {Papaloizou}}{2000}]{nelson00}
{Nelson} R.~P.,  {Papaloizou} J.~C.~B.,  2000, MNRAS, 315, 570

\bibitem[\protect\citeauthoryear{{Ogilvie}}{{Ogilvie}}{1999}]{ogilvie99}
{Ogilvie} G.~I.,  1999, MNRAS, 304, 557

\bibitem[\protect\citeauthoryear{{Ogilvie}}{{Ogilvie}}{2000}]{ogilvie00}
{Ogilvie} G.~I.,  2000, MNRAS, 317, 607

\bibitem[\protect\citeauthoryear{{Ogilvie} \& {Dubus}}{{Ogilvie} \&
  {Dubus}}{2001}]{ogilviedubus}
{Ogilvie} G.~I.,  {Dubus} G.,  2001, MNRAS, 320, 485

\bibitem[\protect\citeauthoryear{{Papaloizou} \& {Lin}}{{Papaloizou} \&
  {Lin}}{1995}]{paplin95}
{Papaloizou} J.~C.~B.,  {Lin} D.~N.~C.,  1995, ApJ, 438, 841

\bibitem[\protect\citeauthoryear{{Papaloizou} \& {Pringle}}{{Papaloizou} \&
  {Pringle}}{1983}]{pappringle83}
{Papaloizou} J.~C.~B.,  {Pringle} J.~E.,  1983, MNRAS, 202, 1181

\bibitem[\protect\citeauthoryear{{Papaloizou}, {Terquem} \& {Lin}}{{Papaloizou}
  et~al.}{1998}]{papaloizou98}
{Papaloizou} J.~C.~B.,  {Terquem} C.,    {Lin} D.~N.~C.,  1998, ApJ, 497, 212

\bibitem[\protect\citeauthoryear{{Pessah}, {Chan} \& {Psaltis}}{{Pessah}
  et~al.}{2007}]{pessah07}
{Pessah} M.~E.,  {Chan} C.~K.,    {Psaltis} D.,  2007, \texttt
  astro-ph/0705.0352

\bibitem[\protect\citeauthoryear{{Pringle}}{{Pringle}}{1992}]{pringle92}
{Pringle} J.~E.,  1992, MNRAS, 258, 811

\bibitem[\protect\citeauthoryear{{Pringle}}{{Pringle}}{1996}]{pringle96}
{Pringle} J.~E.,  1996, MNRAS, 281, 357

\bibitem[\protect\citeauthoryear{{Scheuer} \& {Feiler}}{{Scheuer} \&
  {Feiler}}{1996}]{scheuer96}
{Scheuer} P.~A.~G.,  {Feiler} R.,  1996, MNRAS, 282, 291

\bibitem[\protect\citeauthoryear{{Torkelsson}, {Ogilvie}, {Brandenburg},
  {Pringle}, {Nordlund} \& {Stein}}{{Torkelsson} et~al.}{2000}]{torkelsson00}
{Torkelsson} U.,  {Ogilvie} G.~I.,  {Brandenburg} A.,  {Pringle} J.~E.,
  {Nordlund} {\AA}.,    {Stein} R.~F.,  2000, MNRAS, 318, 47

\bibitem[\protect\citeauthoryear{{Wijers} \& {Pringle}}{{Wijers} \&
  {Pringle}}{1999}]{wijers99}
{Wijers} R.~A.~M.~J.,  {Pringle} J.~E.,  1999, MNRAS, 308, 207

\end{thebibliography}

\end{document}